\documentclass{bmcart}

\usepackage[utf8]{inputenc} 

\usepackage{booktabs} 
\usepackage{placeins}
\usepackage{graphicx}
\usepackage{algorithm}
\usepackage{algorithmic}
\usepackage{multirow}
\usepackage{array}
\usepackage{url}
\usepackage{soul}

\newif\iffigures
\figurestrue 

\iffigures
\else

	\def\includegraphics{}
\fi

\startlocaldefs
\endlocaldefs

\begin{document}

\begin{frontmatter}

\begin{fmbox}
\dochead{Research}

\title{The academic wanderer: structure of collaboration network and relation with research performance}

\author[
   addressref={aff1},                   
   email={p.paraskevopoulos@iit.cnr.it}   
]{\inits{PP}\fnm{Pavlos} \snm{Paraskevopoulos}}
\author[
   addressref={aff1},
   email={c.boldrini@iit.cnr.it}
]{\inits{CB}\fnm{Chiara} \snm{Boldrini}}
\author[
addressref={aff1},
email={a.passarella@iit.cnr.it}
]{\inits{AP}\fnm{Andrea} \snm{Passarella}}
\author[
addressref={aff1},
email={m.conti@iit.cnr.it}
]{\inits{MC}\fnm{Marco} \snm{Conti}}


\address[id=aff1]{
  \orgname{IIT, National Research Council}, 
  \street{Via G. Moruzzi 1},                     %
  \postcode{56124}                                
  \city{Pisa},                              
  \cny{Italy}                                    
}

\end{fmbox}

\begin{abstractbox}

\begin{abstract}
Thanks to the widespread availability of large-scale datasets on scholarly outputs, science itself has come under the microscope with the aim of capturing a quantitative understanding of its workings. In this study, we leverage well-established cognitive models coming from anthropology in order to characterise the personal network of collaborations between scientists, i.e., the network considered from the standpoint of each individual researcher (referred to as ego network), in terms of the cognitive investment they devote to the different collaborations. Building upon these models,  we study the interplay between the structure of academic collaborations, academic performance, and academic mobility at different career stages. We take into account both purely academic mobility (i.e., the number of affiliation changes) and geographical mobility (i.e., physical relocations to different countries). For our investigation, we rely on a dataset comprising the geo-referenced publications of a group of 81,500 authors extracted from Scopus, one of the biggest repositories of academic knowledge.  Our main finding is that there is a clear correlation between the structure of co-authorship ego networks and academic performance indices: the more one publishes and the higher their impact, the larger their collaboration network. However, we observe a capacity bound effect, whereby, beyond a certain point, higher performances become increasingly less correlated with large collaboration networks. We also find that international academic migrants are better at growing their networks than researchers that only migrate within the same country, but the latter seem to be better in exploiting their collaboration to achieve higher impact. High academic mobility does not appear to translate into better academic performance or larger collaboration networks. This shows a different finding with respect to related literature, where scientific productivity is seen as directly linked to mobility. Our results show that, when looking at \emph{impact} of research, this is not necessarily the case.
\end{abstract} 


\end{abstractbox}

\end{frontmatter}

\section{Introduction}
\label{intro}


Science of science~\cite{Fortunato2018} is a recent discipline that aims to get a quantitative large-scale understanding of the ins and outs of the scientific endeavour, leveraging the unprecedented availability of digital data on scholarly output. 
%
Science of science touches upon several issues, such as the tension between innovation (high-risk, high-gain) and tradition~\cite{Foster2015}, citation dynamics~\cite{Eom2011}, gender inequalities~\cite{Moss-Racusin2012}, the impact of teamwork~\cite{Wuchty2007}. The most relevant to this work is the topic of career dynamics, which entails studying how scientific careers unfold over the years, often in response to milestone events like promotions or a move to different institution. Scientific mobility can, indeed, yield major career opportunities for a scientist, like access to better research facilities or collaborations with prestigious colleagues. For this reason, it has attracted a lot of attention from the research community~\cite{deville2014career,sugimoto2017scientists,petersen2018multiscale}. At the same time, each movement may affect the network of already-established collaborations between co-authors~\cite{petersen2018multiscale} (adding new ones or churning old collaborations). Co-authorship partnerships have been shown to play a significant role in a scientistist's career, in special cases (referred to as super ties) contributing to above-average productivity and a 17\% citation increase per publication~\cite{Petersen2015}. For this reason, they are a very important dimension to investigate.

Since academic mobility has the potential to disrupt academic partnerships, as geographic distance and different academic incentives at different institutions may make collaborating harder, in this work we study the effects of academic mobility on co-authorship collaborations. From the mobility standpoint, we consider both pure academic mobility (i.e., a change in affiliation) and geographical mobility (e.g., a relocation to a different country). For studying the collaboration networks of scientists, we leverage results from anthropology, and specifically the well-studied cognitive model known as \emph{social brain hypothesis}. The main idea behind the model is that our brain is allocated a certain capacity for social interactions and that this capacity is distributed unevenly across our social relationships. In particular, we are able to interact \emph{meaningfully} (i.e., to engage in nurturing a relationship at least at the level of sending birthday/Christmas wishes once a year~\cite{hill2003social}) with just a handful of our overall social relationships. This number, whose average is around 150~\cite{dunbar1998social}, is widely known as Dunbar's number. While traditionally associated with purely social contexts, Dunbar's number has been shown to emerge also in many types of organizations~\cite{dunbar2010many}. 
%
Dunbar's theories also characterise how our cognitive efforts are allocated to these 150 peers (typically referred to as \emph{alters}). Specifically, these 150 relationships are structured in concentric groups (known as \emph{circles}) around each individual, and the sizes of these groups are multiples of three. In fact, the typical sizes of the circles are 1.5, 5, 15, 50, 150. The innermost circles contain the strongest relationships, while the opposite holds for the outermost ones~\cite{hill2003social,Zhou2005}. Notably, this hierarchical social structure is known to impact significantly on who we trust~\cite{sutcliffe2015modelling}, on the way information spreads in social networks~\cite{arnaboldi2014information}, and on the diversity of information that can be acquired by users~\cite{aral2011diversity}. This layered social structure is typically represented using an ego network graph~\cite{hill2003social,everett2005ego,lin2001social,mccarty2002structure}, in which the individual, referred to as \emph{ego}, is at the center of the graph, and the edges connect her to the peers (alters) with which she interacts. Figure~\ref{fig:egonet} illustrates the classical ego network structure. 
Interestingly, this ego network structure has been confirmed also for co-authorship relationships~\cite{arnaboldi2016analysis} (please refer to Section~\ref{sec:egonets} for details). 


Building upon these results, in this work we set out to study the interplay between co-authorship ego networks, academic performance, and academic mobility. To the best of our knowledge, this problem has not been considered before in the related literature. For our study, we leverage a dataset extracted from the Scopus repository comprising all the publications of 81,500 scholars at different career stages (ranging from young researchers to distinguished professors). We characterise each scientist based on their academic performance (scientific productivity and impact) as well as their mobility profile (in terms, e.g., of number of institutions changed during the career and domestic vs international mobility). Then, we investigate how these different dimensions of a researcher's career correlate with the structure of their collaboration networks (modelled as ego networks). Specifically, we tackle the following research questions:
\begin{description}
\item [RQ1:] Is the \emph{productivity} of a researcher associated with larger co-authorship circles in their ego network?
\item [RQ2:] Does the \emph{impact} of a researcher translate into larger co-authorship circles?
\item [RQ3:] Is the \emph{mobility (international/domestic)} of an author associated with larger collaboration circles? And how is it related to academic performance?
\item [RQ4:] Does any cognitive capacity bound, similar to those existing in social relationships, manifest also in terms of collaboration network structure?
\end{description}
Our main findings are the following:

\begin{itemize}
    \item In general, there is a \emph{clear positive correlation between sizes of layers and performance indices} (both productivity and impact): the more one publishes or the higher their impact, the larger the layers of their ego network. However, the correlation is particularly high only for the external layers and this is in agreement with the theory on the different cognitive effort that is allocated to the collaborations in the different circles. The outermost circles only require low effort, while the innermost ones necessitate of intense nurturing, hence cannot be grown at will.
    \item The existence of a \emph{capacity bound effect} on the size seems confirmed: beyond a certain point (in terms of performance) higher performances are less and less correlated with large collaboration networks. This confirms at a higher level of detail the similarity between cognitive constraints in general social networks and in scientific collaborations. Also, this result seems to support the findings in~\cite{Petersen2015} about stable partnerships as a strategy for success.
    \item \emph{International and domestic migrants show different efficiency levels in exploiting their collaboration networks}, respectively for productivity and impact. While international migrants seem more effective in exploiting larger networks for higher number of publications, domestic migrants seem to be more effective in exploiting them for achieving high impact. This suggests that international migrations (which typically require more effort on the side of the migrating person) are not always particularly effective in terms of impactful research.
    \item International migrants enjoy a \emph{boost in their co-authorship ego networks earlier in the career} with respect to domestic migrants. The latter, however, \emph{are able to catch up} and end their career at comparatively the same ego network levels.
	\item	 The \emph{most mobile authors} do not publish more and increase the number of their collaborations. Moreover, they are \emph{not able to capitalise mobility in terms of impact}, as for them having bigger ego network sizes is less correlated with high impact with respect to regular authors.
\end{itemize}

The rest of the paper is organized as follows. 
In Section~\ref{sec:related} we present the related work. In Section~\ref{sec:egonets}, we introduce the theory behind the concept of ego networks and we discuss why it is relevant for co-authorship networks.
In Section~\ref{sec:dataset} we overview the available repositories of academic knowledge and we discuss why the Scopus one, used in this paper, offers a clear advantage. We also explain how we collect and pre-process the data. 
In Section~\ref{sec:methodology} we introduce the concept of researcher profile and we present the methodology used for analysing the data. 
Our findings are then discussed in Section~\ref{sec:results}. 
Finally, Section~\ref{conclusion} concludes the paper.

\section{Related Work}
\label{sec:related}




Many studies have focused on the analysis of the effects of academic migrations and the variety of parameters that affect them. 
Moed \etal~\cite{moed2013studying} study the migration flows from a country to another and how the language similarities affect them. 
Five countries (Germany, Italy, the Netherlands, UK, and USA) are considered, and authors are observed for a period of ten years. 
Moed~\etal{} observe that increased international migration corresponds to increased international collaboration, but that the two processes are somewhat distinct. They also show that language similarity is one of the most important factors in relocation decisions. This is one of the few works in the literature that leverages Scopus data for the analysis of scientific mobility, like we do in this paper. Urbinati~\etal{}~\cite{urbinati2019hubs} aim at identifying the countries that play a central role in the international migrations of researchers. Recently, Miranda-Gonzalez~\etal{}~\cite{gonzalez2020scholarly} have used the Scopus data to study internal scholarly migrations in Mexico.


Petersen~\cite{petersen2018multiscale} aims at measuring shifts in researcher profiles before and after a cross-border mobility event. For this study, the author uses the APS Research Review dataset, comprising information about 237,038 physicists and 425,369 scientific papers (filtered such that only 26,170 researchers matched the target career longevity and productivity criteria).
Petersen's focus is on geographic displacements and their effect on the number of citations and collaboration network. 
The take-home messages of this work are that (i) migration is associated with a significant churning in the collaboration network, up to the point where all former collaborations are curtailed (this happens only in 11\% of cases), (ii) the professional ties created after a migration event are less strong hence people will tend to move more also in the future, (iii) career benefits of mobility (measured in terms of number of citations and diversity of research topics and collaborations) are common to all ranks, not just elite scientists. 
Relying on data coming from the Web of Science database, Sugimoto~\etal~\cite{sugimoto2017scientists} set out to study the benefits of global researchers mobility. They found that only a small portion of researchers in their dataset had more than one affiliation. And yet, mobile scholars always achieve higher citation rates than non-mobile ones. Researchers that do move do not severe their ties with their country of origin but instead work as bridges between different countries,  keeping connections to their home country and sometimes, at a point, going back to their country of origin. However, Sugimoto~\etal{} focus on authors that published their first paper after 2008, hence implicitly targeting only young researchers.
Franzoni~\etal{} in~\cite{franzoni2014mover} focus on the effects of academic mobility on research performance (measured in terms of the impact factor of the papers). 
This study focuses on 4 specific research fields and leverages data from 16 countries. 
They highlight the positive effects on the career for researchers that migrate. This work relies on surveys data comprising 14,299 scientists.

Researcher mobility is considered in~\cite{deville2014career,james2018prediction,vaccario2019mobility}, but from a different perspective. James~\etal{}~\cite{james2018prediction} tackle the problem of predicting the next career move of a researcher. Vaccario~\etal{}~\cite{vaccario2019mobility} reconstruct the temporal graph of worldwide movements of scientists and study its properties.
Deville~\etal~\cite{deville2014career} measure the impact of a movement to a higher- or a lower-ranked institution. 
The authors find that a movement to a lower-ranked institution is going to probably affect negatively the profile of a researcher, however the opposite does not hold, with the impact of authors that move to highly-ranked institutions remaining the same. 
%

The geographical composition of the collaboration networks is considered in~\cite{Abramo2011,Kato2013,panzarasa2008patterns}. The main research question in this body of work is whether international collaborations are associated with better research performance with respect to national ones. A positive correlation is found by Abramo~\etal{}~\cite{Abramo2011} and confirmed by Kato and Ando~\cite{Kato2013}. Panzarasa and Opsahl~\cite{panzarasa2008patterns} find that geographic distance among collaborators has positive effects on research performance (but the main focus of their work is the interplay between multi-authorship and specialization). 

The goal of our contribution is to study the interplay among co-authorship ego networks, academic performance, and academic migrations. In contrast to \cite{moed2013studying,urbinati2019hubs,gonzalez2020scholarly}, we are not interested in scholarly migration flows from the countries perspective. Nor do we focus on the ranking and reputation of institutions (as in~\cite{deville2014career}) or on the prediction of the next career move~\cite{james2018prediction}. Differently from \cite{Abramo2011,Kato2013,panzarasa2008patterns}, we are interested in the migration of the ``ego'' author, not in the geographical affiliation of its collaborators (or alters, in ego network terminology). Closest to our work are~\cite{sugimoto2017scientists,petersen2018multiscale,franzoni2014mover}, since their main focus is on individual researchers and the relationship between their academic migrations with their scholarly performance.  
Differently from previous studies, though, we study coauthorship networks through the lens of ego networks (which provide a well-established framework for capturing the different relationship intensity), and we analyse the effect of both cross-border and domestic academic mobility on the structure of researchers' co-authorship ego networks. Furthermore, we split the analysis in six career groups, based on the duration of the career of each researcher. Finally, we separately analyze exceptional cases such as researchers that tend to publish a lot or that have an exceptionally high number of collaborations or exceptionally high mobility, to distinguish their behaviour from that of the ``average'' researcher at the same career stage.

Please note that this paper builds upon our previous work in~\cite{arnaboldi2016analysis} but it is not an extension of it. In~\cite{arnaboldi2016analysis} we have validated the presence of ego network structures (see Section~\ref{sec:egonets} for more details on the work carried out in~\cite{arnaboldi2016analysis}) in co-authorship networks. Here we provide an in-depth analysis of how such structures change depending on the career stage, productivity, impact, and mobility of each researcher. A preliminary analysis of the effect of mobility on the ego networks of researchers has been recently presented in~\cite{paraskevopoulosSocInfo}, where we have investigated the variation of the size and composition of scholarly ego networks before and after a movement. In that work, though, neither the layered structure of ego networks or the different types of migrations have been studied. The results in~\cite{paraskevopoulosSocInfo} can be considered as a preliminary dynamic analysis of how ego networks change upon movements, while the present work is an investigation of the static ego networks across the whole career. Thus, the two works provide distinct but complementary views on co-authorship ego networks.

\section{(Ego) networks of collaborations}
\label{sec:egonets}

Each co-authored paper marks a milestone in a collaboration relationship. When a paper is out in the academic world, its authors have already invested a lot of their time and mental resources to its development. While it is well accepted that academic migrations foster new collaborations and novel results, there might be a price to pay in terms of old collaborations. Relationships may also be affected by geographical distance. Difficulty in interactions due to lack of face-to-face opportunities and time zone difference may also affect the collaboration network of more mobile researchers differently than for less mobile ones. Since the local network of collaborations features similar properties as the purely social one (as shown in~\cite{arnaboldi2016analysis}), it is reasonable to expect that its dynamics may be similar as well. Specifically, findings from anthropology state that social relationships have a maintenance cost~\cite{Roberts2015} and that humans can invest in a defined personal relationship with a limited number of individuals. As a result, as more relationships are added, the intimacy with old relationships inevitably decreases. 

The availability of large-scale longitudinal data about researchers mobility and their collaborations offers a unique playground for studying the effect of mobility on collaboration relationships. In previous work, co-authorship has been treated as a binary relationship between a pair of researchers, without considering the intensity of the collaboration. However, considering the intensity of collaborations may provide novel insights into how researchers structure their work-related network. At the same time, considering collaboration intensity allows us to study the effect of academic mobility on such networks. 
In the remaining of the section, we provide a brief overview of the main results from the social sciences (Section~\ref{sec:relwork_egonetworks}) and we discuss the ego network model for co-authorship networks (Section~\ref{sec:egonets_scientists}). 

\subsection{Background}
\label{sec:relwork_egonetworks}

According to the \emph{social brain hypothesis} from anthropology~\cite{dunbar1998social}, the social life of primates is constrained by the size of their neocortex. Specifically, for humans, the typical group size is estimated around an average of 150 members, a limit that goes under the name of Dunbar's number. 
This limit is related to the cognitive capacity that humans are able to allocate to nurturing their social relationships, and it originates from the time and efforts that primates could devote to grooming. 
The 150 social relationships do not include acquaintances, but only people with which a coherent quality relationship is entertained. In a modern society, this entails, at the very least, exchanging birthday or holiday cards every year~\cite{hill2003social}. 

Unsurprisingly, humans are not fair in how they distribute their cognitive attentions among the 150 important persons they have around them. To the contrary, it is possible to group our social relationships in circles of increasing intimacy, as illustrated in Figure~\ref{fig:egonet}. 
This layered social structure is typically referred to as ego network~\cite{hill2003social,everett2005ego,lin2001social,mccarty2002structure}. The individual (\emph{ego}) is at the center of the graph, and the edges connect her to the peers (called \emph{alters}) with which she interacts. 
Typically, each of us has at least four circles of intimacy~\cite{hill2003social,Zhou2005}, with the innermost one (\emph{support clique}) including our close family and the outermost one \emph{active network} made up of all the people one meaningfully interacts at least once a year.
The circles are conventionally concentric, with the inner layers contained in the outer ones. Many people also feature an additional inner layer (contained in the support clique) comprising on average $1.5$ alters for which they have a very high emotional investment~\cite{Dunbar2015}. 
The ego-alter tie strength is typically computed as a function of the frequency of interactions between the ego and the alter. The ego network structure is characterised by a striking regularity, with an approximately constant ratio of 3 between the size of consecutive layers. 


The social brain theory was developed for the offline world, in which relationships were nurtured via face-to-face interactions, letters, or landline phone calls. 
However, Dunbar's model was confirmed to hold also on several non-face-to-face communication means (such as email communications~\cite{Haerter2012}, mobile phone calls~\cite{Miritello2013}, and Online Social Networks~\cite{Dunbar2015,gonccalves2011modeling}). 


\subsection{Ego networks of scientists}
\label{sec:egonets_scientists}

More interestingly with respect to the scope of this paper, many types of social organizations are arranged around the layered structure discussed in the previous section. This is the case of, among others, modern military organizations~\cite{Zhou2005} and investors networks~\cite{johansen1999predicting}. In a previous work~\cite{arnaboldi2016analysis}, we have confirmed the existence of this layered structure for academic collaborations as well. Our goal here is not to validate the presence of similar layered structures for co-authorship networks, as this has been already done in~\cite{arnaboldi2016analysis}. 
Our goal is to understand how this structure is related to the academic performance and the mobility profile of researchers, especially for what it concerns the size of the layers. To this aim, we recall below the reference ego network model for academic collaborations defined in~\cite{arnaboldi2016analysis}. 

We consider the researcher \textit{i} as the ego and all the co-authors appearing in the research publications of the ego as the alters. From all the publications on which the ego \textit{i} and the alter \textit{j} have worked together, the intensity (i.e., tie strength) $t_{ij}$ of the collaboration between $i$ and $j$ can be computed as follows:
\begin{equation}
    t_{ij} = \frac{1}{d_{ij}} \sum_{p \in \{P(i) \cap P(j)\}} \frac{1}{k(p)-1},
    \label{eq:egonet_compute}
\end{equation}
where $d_{ij}$ represents the duration of the co-authorship relation (i.e., time between first co-authored paper and last paper published by the ego \textit{i}), $k(p)$ denotes the number of authors of paper $p$, and $P(i)$ denotes the set of papers authored by researcher $i$. As a filtering parameter, in order to have a sufficiently long observation period for each collaboration, we consider as adequate only the co-authorship relations longer than six months, thus discarding those that have been initiated the last six months of the ego's career. We have tested thresholds at both at six months and twelve months. The difference, on this specific dataset, was negligible, hence we opted for the 6-month threshold in analogy with~\cite{arnaboldi2016analysis}.

Each co-authored paper refreshes and nurtures the relationship, hence it adds a quantum of tie strength to the collaboration. 
This quantum is inversely proportional to the number of co-authors, thus this definition of tie strength embodies the notion that a smaller number of co-authors typically implies stronger collaborations. Note that the tie strength in Equation~\ref{eq:egonet_compute} is basically a \emph{frequency} of common co-authorship between $i$ and $j$ (modified according to the number of co-authors), as we divide by the length of time $d_{ij}$ since the co-authorship has been established. The latter implies that this tie strength definition factors in the ups and downs of collaborations, as it decreases if no new co-authored publications are produced over time. Analyses leveraging this tie strength definition are usually referred to as \emph{static}, because this definition yields an average picture of the state of the collaboration for its entire duration.  The complementary analysis, referred to as dynamic~\cite{Arnaboldi2013b}, entails studying the variation of the tie strength over time. This approach would be suitable, for example, to investigate whether the academic status of a collaboration affects the future evolution of the co-authorship ego network, or how the overall collaboration network growth process affects individual ego networks over time.  
Investigating such dynamic aspects for scholarly collaboration is left as future work.
In this work, we focus on the static analysis, which has been shown to be very effective in identifying social cognitive constraints in the related literature (see Section~\ref{sec:relwork_egonetworks}).
Finally, note that the frequency of interaction is one of the typical measures of tie strength used in the literature~\cite{Dunbar2015}. This definition also implies that we are working with \emph{weighted} networks. 

Having computed the tie-strength between the ego and the alters, we cluster the co-authors based on the tie-strength, in order to obtain the circles of collaborations.
 Thus, we cluster tie strengths between the ego and the alters in 5 groups.
For getting the optimal break points of each cluster, we use the Jenks natural breaks classification method~\cite{Jenks1977}.  
As we will see in Section~\ref{sec:results}, five is typically the optimal number of circles for our dataset (this is in line with previous results, as discussed in Section~\ref{sec:egonets}). 
Once we confirmed this, we adopt the common practice in the related literature~\cite{arnaboldi2016analysis}: we map all the egos’ collaborations into 5 circles. This practice reduces the complexity of the analysis, because it allows us to analyse all the egos together, rather than splitting them based on the optimal number of collaboration circles of each ego.
For getting the optimal break points of the fixed 5 clusters, we use the Jenks natural breaks clustering method~\cite{Jenks1977}. 
As a result, for each ego we obtain an ego network of 5 layers, where the first one contains the strongest collaborations and the fifth contains the weakest ones. The ego network size is defined as the total number of collaborations existing across all the ego network layers.

%

\section{The dataset}
\label{sec:dataset}

In this section, we discuss why we chose Scopus as our reference dataset (Section~\ref{sec:data_landscape}), how we collect our data (Section~\ref{sec:data_collection}), and how we preprocess and filter it (Sections~\ref{sec:mobility_extraction}-\ref{sec:data_filtering}). 
 
\subsection{The academic data landscape}
\label{sec:data_landscape}

Finding a suitable dataset for studying academic mobility entails obtaining longitudinal information on the affiliations of a reference set of authors. 
Discarding unscalable approaches that involve surveys~\cite{Wang2019} or manually checking the website/CVs of the authors in order to compile the affiliation history, we can rely on several platforms that provide information about individual authors (Table~\ref{tab:repositories}). 
Some of them export free or premium APIs that can be used to retrieve information, some others offer dumps of their data base covering a specific time span or a certain fraction of authors. 
When it comes to listing affiliations (which, as explained in Section~\ref{sec:mobility_extraction}, we use as proxy for detecting mobility), some of them (like DBLP, used in~\cite{Servia-Rodriguez2015}) do not provide the information at all. 
Google Scholar (used in~\cite{arnaboldi2016analysis,Servia-Rodriguez2015}) lists only the most recent affiliation if this information is provided by the authors themselves. 
In ORCID, researchers can enter their full affiliation history. 
The resulting affiliations are typically reliable and accurate, but the drawback of this approach is that, for most author profiles, the list of affiliations is often missing or incomplete, as it is provided on a voluntary basis. 
ORCID data has been used in~\cite{urbinati2019hubs,Bohannon2017} for studying researchers mobility. 
A more scalable approach is based on the automatic extraction of the affiliation directly from papers and their metadata. 
Pursuing this strategy, repositories like APS, MedLine/PubMed (since 2014), Microsoft Academic, Scopus, and Web of Science (WoS) are able to provide affiliation information for each paper and for each author in their database. 
WoS, used in~\cite{sugimoto2017scientists} for studying the global circulation of scholars, requires premium fees for accessing affiliation information. 
Hence, we do not consider it further in this study.
APS has been often used as reference repository in the related literature~\cite{deville2014career,petersen2018multiscale,james2018prediction,Wang2013,Qi2017}. 
MedLine was used in~\cite{vaccario2019mobility}. 
However, while the affiliation per se is useful in studying academic mobility (i.e., movements from one institution to another one, as in~\cite{deville2014career}), it needs to be further processed if the goal is to study geographic mobility as well. 
For example, Petersen~\cite{petersen2018multiscale} extracts geographical information from APS data by matching country names and codes in the affiliation text string (geoparsing). 
Another approach entails querying location services (such as Google Maps) in order to retrieve the position of the university a scientist is affiliated to~\cite{james2018prediction}. 
Much more convenient is to rely on databases that already provide preprocessed geographical information.  
The Scopus API does that, offering geo-referenced information about the affiliation up to the city level granularity. 
For this reason, in this work we focus on affiliation data coming from the Scopus repository.

\begin{table}[t]
\caption{Summary of academic knowledge/career repositories.}
\label{tab:repositories}
\begin{tabular}{@{}l>{\raggedright}p{0.14\textwidth}>{\raggedright}p{0.1\textwidth}>{\raggedright}p{0.16\linewidth}l l@{}}
\toprule
\textbf{Repository} & \textbf{Access} & \textbf{API} & \textbf{Aff. History} & \textbf{Georefe.} & \textbf{Papers} \\ \midrule
Google Scholar & - & No (scraped) & No & No & \cite{arnaboldi2016analysis,Servia-Rodriguez2015} \\
DBLP  & Free & Yes & No & No & \cite{Servia-Rodriguez2015}\\
APS  & Free & No (dump) & Yes & No & \cite{deville2014career,petersen2018multiscale,james2018prediction,Wang2013,deville2014career,Qi2017}\\ 
WoS  & Free/Premium & Yes & Yes (Premium) & No & \cite{sugimoto2017scientists,Abramo2011,Kato2013}\\
MedLine/PubMed  & Free & Yes & Yes & No & \cite{vaccario2019mobility}\\
Forsythe-CRA  & Free (hand-curated, North America faculty only) & No & Yes & No & \cite{Way2017,Way2019}\\
Microsoft Academic  & Free & Dump & Yes & No & \cite{Pramanik2019}\\
ORCID  & Free/Premium & Yes &  Yes & No & \cite{urbinati2019hubs,Bohannon2017}\\
Scopus & Free for Scopus-subscribing institutions & Yes  & Yes & Yes & \cite{moed2013studying,gonzalez2020scholarly,paraskevopoulosSocInfo} \\
\bottomrule
\end{tabular}
\end{table}


\subsection{Data collection}
\label{sec:data_collection}

In order to identify a set of authors to study, we focus on the group of authors whose profiles and publications were collected by Arnaboldi~\etal{} in~\cite{arnaboldi2016analysis}. Arnaboldi~\etal{} crawled the authors' profiles in Google Scholar starting from a single category (``computer science''), then accessed all the categories found in the visited profiles, iterating the procedure until no new categories were found. Since Google Scholar does not provide the history of affiliations, nor their geographical position, the dataset collected by Arnaboldi~\etal{} cannot be used directly for our analysis. 
Thus, in order to leverage the georeferenced affiliation history provided by Scopus, we downloaded afresh from Scopus all the publications authored by the Google Scholar researchers studied in~\cite{arnaboldi2016analysis}. The download took place in November 2019.
We used the official Scopus APIs\footnote{version: 2.0.1 (currently renamed to ``pybliometrics"~\cite{rose2019pybliometrics:-scriptable}) - \url{https://dev.elsevier.com/scopus.html}}. Specifically, we used the 
Author Search API, the Author Retrieval API, and the Affiliation Retrieval API, as described below.  

Since an automatic mapping between the researcher IDs across the two platforms (Google Scholar and Scopus) does not exist, we established this mapping ourselves. %
Given that two researchers can share the same name, it is not enough to match the first and last name in the two datasets. 
In order to resolve this ambiguity, we decided to match each Google Scholar researcher~$r_{g}$ with a Scopus researcher~$r_{s}$ of the same name that has authored at least one paper with the same title as those authored by $r_{g}$. 
In practical terms, we passed the name and surname of the researcher~$r_{g}$ to the Scopus Author Search API, retrieving all the researchers in Scopus matching the (string) name passed. This allowed us to get the IDs of the \emph{potential matches} $r_s$ in the Scopus database, while also getting their full publication record. Through these publication records, we were able to match the titles of $r_{g}$'s papers in the Scholar database with those of the potential matches in the Scopus database. We establish a match between $r_{g}$ and a potential match $r_s$ if $r_s$ shares with $r_g$ at least a publication with the same name. Then, we use the Scopus ID of $r_s$ to get, via the Author Retrieval API, the full author profile: full name, h-index, number of publications, citations received and research areas. The full history of affiliation is obtained from the publication records. In case of incomplete affiliation information in the publication record, we queried the Affiliation Retrieval API.

In the end, we successfully identified 102,236 authors (out of the initial 285,577 in the Google Scholar dataset) that have authored 5,847,723 publications. For these authors, we have their ID in the Scopus database, the total number of the citations received and their h-index (at the time we collected the data), as well as the complete list of their publications (geo-referenced, timestamped, with all co-authors). 
Please note that this Scholar-Scopus mapping does not directly affect the ego networks of authors. In fact, it simply reduces the pools of egos, since the Scholar ego authors for which a match is not found in Scopus are dropped. The mapping is not applied to co-authors (alters), so they are not affected. However, the ego networks extracted from Google Scholar may be different from those extracted from Scopus, since the papers published on the two platforms do not completely overlap.


\subsection{Extraction of academic mobility profiles}
\label{sec:mobility_extraction}

The raw Scopus dataset yields a list of geo-referenced affiliations per author, timestamped with the publication date of each paper. 
%
Since the actual movements of a researcher cannot be directly tracked from a scholarly database, we need to use the affiliated institutions listed in papers as a proxy of the locations visited by a researcher. 
This approach is widespread in the related literature~\cite{sugimoto2017scientists,petersen2018multiscale,james2018prediction}, since it allows for large-scale automated collection of academic movements. 
However, there are several limitations. First, there is always a time lag between a change in affiliation and the first published paper that allows us to detect it. 
Second, a change in affiliation does not always imply a physical relocation to a new institution, especially for more senior researchers. 
Third, the detection and proper management of individual multiple affiliations is not trivial. On the one hand, the multiple affiliations tend to be quite noisy in the paper metadata, as there is no standard way for reporting them and this makes their extraction more complex. On the other hand, since a person cannot be in two geographical locations at exactly the same time, an author with multiple affiliations has to be assigned to a single one of them, according to a certain criterion (see below for a detailed discussion on the one we have chosen).
Despite these limitations, at the moment, to the best of our knowledge, inferring mobility from the affiliation history in academic repositories is the only viable alternative to surveys, which are costly and do not allow researchers to obtain comparable large-scale datasets. 

In order to reconstruct the academic trajectory of the researchers in our dataset, we create, for each of them, a time series where, for each point, the x-coordinate is given by the month and year (timeslot) the researcher has published a paper, and the y-coordinate by the geo-referenced affiliation reported by the researcher in the paper. Recall that each affiliation can provide geographical information at different granularity levels (country, city). When the specific granularity is not important, we will refer unambiguously to them as \emph{locations}.
Two issues may be encountered in the construction of this time series. Specifically, the author may have multiple publications with different affiliations associated with the same timeslot or she may have more than one affiliation per paper. 
In both cases, the most representative affiliation should be kept. 
%
%
We developed a probabilistic model that keeps the affiliation that is referred more frequently in a specific timeslot. 
In case this is not enough to extract the most representative affiliation, we take into consideration the affiliations used in the temporally closest (either in the past or in the future) timeslot. 


Once we have the location time series (each representing an academic and geographical trajectory), we can get the distance between consecutive movements. To this aim, we leverage the Google Maps API, to which we pass the locations of two consecutive timeslots in the time series. However, some affiliations may have missing georeferenced data, such as a missing city or country. In order to bypass this issue, whenever a city does not exist in the affiliation record, we query Google Maps for the address of the institution directly. A drawback that emerged was the mistaken identification of some institutions, a problem that introduced noise in the computation of the distances travelled. However, after manually checking these cases, we found out that the percentage of the mistaken identification was lower than 1\%. Thus, we do not further handled this issue. 

\subsection{Meaningful co-authorships}
\label{sec:data_filtering}

As discussed in Section~\ref{sec:egonets}, each collaboration must correspond to a non-negligible cognitive effort spent for maintaining the relationship. 
This requirement clashes with the presence, in the dataset, of papers with very many authors. 
In that case, it can be hardly argued that those authors have strong personal collaborations between themselves (at least, this is unlikely to be true for all the authors involved). 
%
In order to handle this problem, we focus on the distribution of the number of co-authors per publication. 
The average number of co-authors per paper in our dataset is 26, with a standard deviation of 219. 
The maximum number of co-authors in a paper is 5,563, while 96.8\% of the papers have less than 26 co-authors. 
Setting the latter as the threshold for assuming a meaningful collaboration, we dropped from the analysis the papers that have more than 26 authors (thus keeping approx. 97\% of papers), reducing the number of the usable papers to 5,659,268. 
Our main goal here is to discard the long tail of extremely large collaborations, in the absence of a clear reference threshold from the related literature. 

Finally, from these meaningful co-authorship relations, we construct the ego networks as described in Section~\ref{sec:egonets_scientists}. For a minority of egos (those with a  small optimal number of circles according to Mean Shift), we could not populate the fixed five circles properly. Thus, we do not consider these scientists in our analysis. 
Please note, though, that this does not affect the finding about the layered structure of co-authorship ego networks, as the latter is confirmed (see Section~\ref{sec:exploratory}) before dropping the authors with empty layers.
In the end, the dataset we use for our analysis includes 81,500 authors with complete and geo-referenced ego networks.  



\section{Methodology}
\label{sec:methodology}

The goal of this work is to characterise the interplay between the personal networks of scientists, their mobility, and their academic performance. To this aim we first need to compile a complete researcher profile for each author, based on the data extracted from Scopus. This task is described in Section~\ref{sec:researcher_profile} below. Then, in Section~\ref{sec:correlation_analysis} we discuss our approach for establishing a relationship between the different dimensions of the researcher profiles.

\subsection{Building the researcher profile}
\label{sec:researcher_profile}

Leveraging the dataset described in Section~\ref{sec:dataset}, we obtain, for each researcher, the following set of features (summarised in Table~\ref{tab:features_summary}).

\begin{table}[t]
\begin{center}
\caption{The researcher profile.}
\label{tab:features_summary}
\begin{tabular}{ >{\bfseries}l l l} 
\toprule
Identity & Researcher ID & \\ \midrule
Research areas & see Table~\ref{table:researchareas} & \\  \midrule
\multirow{2}{5em}{Academic performance} & Productivity & Number of publications \\  \cmidrule{2-3}
& Impact  & h-index \\ \midrule
Career stage & see Table~\ref{table:career_samples} & \\ \midrule
\multirow{5}{5em}{Mobility measures} & Number of institutions & \\ \cmidrule{2-3}
 & \multirow{2}{6em}{Number of locations} & City granularity\\ \cmidrule{3-3}
 &  & Country granularity \\ \cmidrule{2-3}
 & \multirow{2}{6em}{Distance travelled} & Average \\ \cmidrule{3-3}
 & & Cumulative \\ \cmidrule{2-3}
 & Migration status & Domestic, international \\
  \midrule
Collaboration network measures & Circle size & For circles C1, ..., C5 \\
\bottomrule 
\end{tabular}
\end{center}
\end{table}

\noindent \textbf{Research areas: }
The research areas a scholar has been actively working on are identified from the 27 master categories shown in Table~\ref{table:researchareas}, which correspond to the ones provided by the Scopus API.

\begin{table}[h]
    \centering
    \caption{Research Areas}
    \label{table:researchareas}
    \begin{tabular}{ll}
        \toprule
        AGRI: Agricultural and Biological Sciences & ARTS: Arts and Humanities \\ 
        BIOC: Biochemistry, Genetics and Molecular Biology & BUSI: Business, Management and Accounting \\ 
        CENG: Chemical Engineering & CHEM: Chemistry  \\ 
        COMP: Computer Science & DECI: Decision Sciences \\ 
        DENT: Dentistry & EART: Earth and Planetary Sciences \\ 
        ECON: Economics, Econometrics and Finance & ENER: Energy \\ 
        ENGI: Engineering & ENVI: Environmental Science \\ 
        HEAL: Health Professions & IMMU: Immunology and Microbiology \\ 
        MATE: Materials Science & MATH: Mathematics \\ 
        MEDI: Medicine & MULT: Multidisciplinary \\ 
        NEUR: Neuroscience & NURS: Nursing \\ 
        PHAR: Pharmacology, Toxicology and Pharmaceutics & PHYS: Physics and Astronomy \\ 
        PSYC: Psychology  & SOCI: Social Sciences \\ 
        VETE: Veterinary & \\ \bottomrule
    \end{tabular}
\end{table}

\noindent \textbf{Academic Performance Features: }
The two indexes of academic performance we consider are the \emph{number of papers} a researcher has published and the \emph{h-index} the researcher has achieved. While the former describes the productivity quantitatively, the latter is a measure of the impact of the work within the research community. 

\noindent \textbf{Career Stage: }
Since publications are accumulated over time, comparing performance features between young and senior researchers is not sensible, as the same numbers that would categorise a young researcher as brilliant would put a senior researcher in the under-performing category. 
Thus, we apply a binning based on the duration of the research career, measured as the time distance between the last and first publication in our dataset. 
Note that studying the different career phases separately also allows us to mitigate the effect of the time scale at which the scientific collaboration networks evolve. 
We consider the categories of: PhD students (0 to 3 years), young researchers (3 to 6 years), assistant professors (6 to 10 years), associate professors (10 to 28 years), full professors (28 to 38 years) and distinguished professors (38+ years). The group definition and their sample size\footnote{The low number of authors classified as PhD students in our dataset is due to the fact that we rely only on authors that were present in the Google Scholar dataset collected by Arnaboldi et al.~\cite{arnaboldi2016analysis} in November 2013.  Since then, the majority of those authors have progressed in their career.} in our dataset can be seen in Table~\ref{table:career_samples}. 
Please note that the career group names are intended to be a proxy for the career length in years, aimed at making the explanations more reader-friendly. In fact, the Scopus dataset  does not allow us to derive the real professional position of individual scholars.

\begin{table}[h]
    \centering
    \caption{Career groups and corresponding sample size (number of researchers in our dataset).}
    \label{table:career_samples}
    \begin{tabular}{lll}
        \toprule
        \textbf{Career Group}             & \textbf{Duration}              & \textbf{Sample Size}   \\ \midrule
        PhD Students             & (0, 3) years        & 15     \\ 
        Young Researchers        & (3, 6) years        & 2,928  \\ 
        Assistant Professors     & (6, 10) years       & 16,112 \\ 
        Associate Professors     & (10, 28) years      & 51,506 \\ 
        Full Professors          & (28, 38) years      & 7,986  \\ 
        Distinguished Professors & \textgreater 38 years & 2,953  \\ \bottomrule 
    \end{tabular}
\end{table}

\noindent \textbf{Academic Mobility Features: }
Academic mobility can be studied at the institution level and at the geographical level. The main barrier to the latter is the granularity of geo-referenced data one can obtain/extract per researcher. As mentioned before, Scopus provides georeferenced affiliation information up to the city granularity. For each author we can, thus, provide: the number of institutions with which they have been affiliated over the course of their career, the number of different affiliation locations (both at the city and the country granularity), as well as the distance (in km) between consecutive affiliations.

A relocation to a new institution may take its toll on a researcher and their collaborations. This is particularly true when relocating implies moving to a different country. The geographical distance, the different academic culture and incentives might make it harder for a researcher to nurture its collaboration network. For this reason, among the mobility features we also measure whether the researcher has moved at least once to a different country or not. We refer to the scientists in the former category as \emph{international migrants}, while the latter are referred to as \emph{domestic migrants}. We acknowledge that this classification may seem loose, since a domestic migration in a very large country may be very different from one within a small country. However, movements within the same country typically enjoy a greater homogeneity of norms and practices with respect to international movements. For this reason, we believe that this classification is well-suited to capture the cognitive effort implied by the two classes of movements. 


\noindent \textbf{Collaboration Network Features: }
Finally, as explained in Section~\ref{sec:egonets_scientists}, we cluster each ego's collaborators in five groups, generating five ``concentric circles'' according to the ego network model. In the following, we refer to the circles as C1, C2, \ldots, C5, from the most intimate one (strongest collaborations) to the least intimate one, respectively. 

\subsection{Correlation analysis}
\label{sec:correlation_analysis}

%
%

In this work we carry out an analysis of the relationship between the different dimensions in the researcher profile, with a particular focus on the co-authorship ego network. The specific relationships we investigate are explained in detail in Section~\ref{sec:results}. In general, we study the correlation between different elements of a researcher's profile, such as, e.g., the correlation between ego network layers size and number of affiliations changed.
For studying how a pair of features $X,Y$ is correlated, we use the Pearson correlation, which can be computed as follows: 
\begin{equation} \label{eq:pearson}
r_{xy} = \frac{\sum_i (x_i - \overline{x})(y_i - \overline{y})}{\sqrt{\sum_i (x_i - \overline{x})^2 } \sqrt{\sum_i (y_i - \overline{y})^2 }},
\end{equation}
for $n$ points $\left\{(x_{1},y_{1}),\ldots ,(x_{n},y_{n})\right\}$. Note that the Pearson correlation detects associations between the $X,Y$, not causal relationships, thus we do not claim any causal effect between mobility and ego network features, just co-occurrence. 
Note also that Pearson correlation does not make any assumption on the underlying distribution of random variables X and Y beyond requiring their variance to be defined~\cite{dekking2005modern}. This is because, intuitively, the correlation is simply a standardized version of the covariance, allowing the latter to become independent of the units in which X and Y are measured.
In Section~\ref{sec:results}, we also study separately the top researchers (from the performance or mobility standpoint) from the others. Correlations are known to be strongly affected when one of the variables is constrained.
In this case, we apply a correction to the Pearson correlation known as Thorndike correction~\cite{sackett2000correction,thorndike1949personnel}. It is well documented that corrected correlations are less biased than uncorrected ones, under a wide range of assumptions violations~\cite{gross1983restriction,holmes1990robustness}. Specifically, we use the Thorndike Case~2 correction when we directly restrict the range of the X variable (Section~\ref{sec:performance_and_ego}). Instead, we use the Thorndike Case~3 correction when we restrict the range of a third variable Z, which however might induce an indirect range restriction on X (Section~\ref{sec:mobility_and_ego}).

The confidence in the reported correlations is measured relying on bootstrapped confidence intervals. This non-parametric approach yields much more robust result with respect to methods (e.g. Fisher transformation~\cite{fisher1915frequency}) that assume X and Y to be approximately normal. Specifically, we use the bias-corrected and accelerated (BCa) bootstrap~\cite{efron1987better}, which adjusts for both bias and skewness in the bootstrap distribution. 
In this work, we use 95\% confidence intervals, obtained performing $10,000$ bootstrap replicas.





\section{Results}
\label{sec:results}

We first present, in Section~\ref{sec:exploratory}, a brief exploratory analysis of the dataset. Then, in Section~\ref{sec:performance_and_ego}, we study how the structure of the co-authorship ego network correlates with the considered performance indices, i.e.,  the number of publications and the h-index of researchers. In Section~\ref{sec:mobility_and_ego} we focus on the correlation between the structure of the co-authorship ego network and the mobility characteristics of researchers (number of institution visited, domestic or international migration). 

\subsection{Researcher profiles overview}
\label{sec:exploratory}

In order to better understand the type of researcher profiles featured in our dataset, we briefly explore the main characteristics of the features discussed in Section~\ref{sec:researcher_profile}.
We start by analysing the research areas (see Table~\ref{table:researchareas}) covered by the authors in our dataset (Figure~\ref{fig:histogram_authorsperarea}), in order to assess the representativeness of our sample. We observe that the most popular research areas never account for more than 8-9\% of the sample, and almost all areas cover at least 1\% of the sample. Hence, we dot not expect the results to be substantially skewed by one ``outlier" subject area. 



\noindent
We now move to analysing the career duration, whose histogram is shown in Figure~\ref{fig:histogram_careerduration}. 
The average career duration is 17.1 years and the standard deviation in our dataset is 9.2. Almost all the career groups have a sample size of some thousands (Table~\ref{table:career_samples}).
The vast majority of the scholars in our sample can be classified as assistant professors or associate professors. 
PhD students are very few, hence they will not be considered further in the analysis. 


Next, we focus on the academic performance indices.
From the impact standpoint (Figure~\ref{fig:histogram_hindex}), the dataset spans a significant range of h-index values, including also some top researchers with h-index $\geq 40$, plus some truly outstanding researchers with h-index above 60. 
The average h-index in the dataset is 17.2 (sd = 13.47), which is consistent with the expected h-index at the most popular  career stages (mostly assistant/associate professors) in the dataset. 
In terms of productivity (Figure~\ref{fig:histogram_publications}), we observe that the average number of published papers is 68 (sd=90.5, median=40), with a long tail of authors publishing up to 2,773 papers in their career.



Let us now have a look at the mobility profile of the researchers in our dataset. 
In terms of affiliations (Figure~\ref{fig:histogram_numberofmovements}), our researchers are quite mobile, with the majority of them having changed between one and ten affiliations during their career. 
The observed average is 7.8 with a standard deviation of 10. 
Approximately 3\% of these scientists can be classified as extremely mobile, having changed more than 25 affiliations in their career. 
We cannot exclude that these numbers are partially inflated by the way the affiliation time series is reconstructed. 
As discussed in Section~\ref{sec:mobility_extraction}, inferring the correct time series of affiliations from the time series of published papers is made difficult by the time lag between different types of publications (e.g., journal vs conference), which may create sequences like $\{$..., affiliation1, affiliation2, affiliation1, affiliation2$\}$ where the second appearance of affiliation1 is only due to a delay in publishing with that affiliation, rather than to an actual return to affiliation1. 
The preprocessing discussed in Section~\ref{sec:data_filtering} was aimed at mitigating the problem but, due to the lack of ground truth, it is impossible to assess its efficacy. 


Recalling that we are interested in distinguishing between domestic and international movements, we now look at the number of international vs domestic migrants. 
We observe a very balanced situation, with 40,749 \textit{international} and 40,751 \textit{domestic} migrants in our dataset. 
This implies that several of the affiliation changes detected in Figure~\ref{fig:histogram_numberofmovements} are within the same country. 
When studying academic migrations, not all movements have the same impact. 
A relocation between two institutions in Europe is not comparable, in terms of personal effort and impact of the established collaboration, with a movement between, e.g., Europe and East Asia. 
Different time zones, large distances, different cultures make the move much more challenging, hence worth being studied separately. 
In Figure~\ref{fig:histogram_distofmovements} we plot the histogram of the average (per author) length of movements travelled by the authors in our dataset. 
While the majority of movements are in relatively close proximity, we observe some long-distance relocations. 
As a reference, the average distance in km between Europe and the US is approximately 8,000 km. 
The average trip-distance travelled by a researcher is 2,560 km, with a standard deviation of 3,200km 




Finally, we study the ego networks of the authors in our dataset. Preliminarily, we verify what is their optimal number of collaboration circles. To this aim, we leverage the methodology discussed in Section~\ref{sec:egonets_scientists}, applying the Mean Shift clustering algorithm to the authors in our dataset. Figure~\ref{fig:histogram_optimalcircles} shows the distribution of the obtained optimal number of circles. This distribution is centered around 5, as it was the case for the distribution of the social circles of purely social interactions~\cite{Dunbar2015}. Thus, without loss of generality, in the following we will restrict ego networks to having 5 layers, which are obtained using the Jenks clustering algorithm~\cite{Jenks1977}. In Table~\ref{table:layer_size} we summarise the properties of the co-authorship ego networks for the scientists in our dataset. Specifically, we report the mean circle size at the different career stages, that we have computed as discussed in Section~\ref{sec:egonets_scientists}, and the scaling ratio, which is defined as the ratio between the sizes of consecutive circles. Recall that the typical sizes of circles in purely social networks are 1.5, 5, 15, 50, 150, with a scaling ratio around 3. These numbers should be treated as reference fingerprints of social ego network structures, with the actual sizes of circles varying, possibly quite significantly, around these reference values, depending on the specific social environment considered. 
One limitation of our analysis (common to all works relying on data automatically extracted from publication repositories) is that we are only able to capture the collaboration strength at a coarse granularity: e.g., we cannot exactly pinpoint the co-authors on which the collaboration efforts hinged. Currently, though, these datasets provide the best large-scale proxies of collaboration strengths available for research.
The results in Table~\ref{table:layer_size} show that, as expected, the ego networks of scientists at the beginning of their career are smaller than those of more established colleagues. Despite these important difference, when focusing on individual career stages, we observe a striking regularity in the way collaborations (hence, cognitive efforts) are distributed across the layers, with a scaling ratio $\sim 2-3$. In order to take into account the effect of the career stage on the size of ego network layers, in the following sections we always assess the impact of the career stage on the specific correlation under study.



\begin{table}[t]
\centering
\caption{Ego network circles at different career stages: mean size and scaling ratio. }
\label{table:layer_size}
\begin{tabular}{c c c c c c c c c c c }
\toprule
\multirow{2}{*}{} & \multicolumn{2}{c}{Young Res} & \multicolumn{2}{c}{Assist Prof} & \multicolumn{2}{c}{Assoc Prof} & \multicolumn{2}{c}{Full Prof} & \multicolumn{2}{c}{Dist Prof} \\ \cmidrule(r){2-3} \cmidrule(r){4-5}  \cmidrule(r){6-7}  \cmidrule(r){8-9} \cmidrule(r){10-11} 
                  & mean           & sr          & mean            & sr           & mean           & sr           & mean           & sr           & mean           & sr            \\ \midrule
1st Circle        & 2.9            & -           & 3.2             & -            & 3.8            & -            & 4.3              & -            & 4.1            & -           \\ 
2nd Circle        & 5.5            & 1.9        & 6.7             & 2.1          & 9.3            & 2.4          & 11.9           & 2.8          & 11.6           & 2.8           \\ 
3rd Circle        & 9.8            & 1.8        & 12.9              & 1.9          & 20.6           & 2.2          & 29           & 2.4          & 28.9           & 2.5          \\ 
4th Circle        & 16.4            & 1.7        & 24.1             & 1.9          & 44.3           & 2.2          & 69.7             & 2.4          & 73           & 2.5          \\ 
5st Circle        & 26.2            & 1.6        & 42.5             & 1.8          & 97.9           & 2.2          & 188.2          & 2.7          & 212          & 2.9         \\ \bottomrule
\end{tabular} 
\end{table}

\subsection{Academic performance and co-authorship ego networks}
\label{sec:performance_and_ego}

After the preliminary characterisation of the researchers in the dataset carried out in the previous section, we now start investigating the relationships between measures of productivity and impact, and the observed ego network structure. 
To this aim, we focus on the correlation between the features of interest, as discussed in Section~\ref{sec:correlation_analysis}. 
Recall that the ordering of ego network layers has an inverse relationship with the collaboration strength: the innermost layers are small and contain only the strongest connections, while the opposite holds for the outermost layers.

\subsubsection{Productivity and circles of collaborations}
\label{sec:productivity_and_ego}

The first question that we tackle is \emph{whether higher productivity is associated with more numerous collaborations in all ego network circles}. On the one hand, it is expected that more productive people  (people that publish a lot of papers) tend to also collaborate more. However, as discussed in Section~\ref{sec:relwork_egonetworks}, in human social relationships each collaboration consumes cognitive efforts on the ego's side. Hence the ego cannot increase her network and invest the same amount of time/efforts as before into each prior relationship. So, while it is easier to grow outermost layers (which include weak relationships), expanding innermost circles is much harder. We conjecture that similar properties may hold also for ego co-authorship networks. We provide the scatterplot of layer size vs productivity for the five circles in Figure~\ref{fig:egonetwork_pubs}, and the corresponding correlation values in the first column of Table~\ref{table:pearson_perLayer}. While the linear trend is evident for the outermost circles, as we move inwards the correlation starts to decrease. When we reach the first circle (Figure~\ref{fig:egonetwork_pubs}) the strong correlation (r=0.80) of C5 has significantly faded away (r=0.26 in C1). In Table~\ref{table:pearson_perLayer}, we also provide the correlation between the career length and productivity, in order to rule out that the strong correlation we observe at the outermost layers is only a side-effect of joint correlation on simpler features of the researcher profile (Table~\ref{tab:features_summary}). Since the correlation between career length and productivity is generally lower than the correlation between productivity and layer size, we can rule out this hypothesis.


\begin{table}[t]
\centering
\caption{Correlation of productivity, impact, and mobility with layer size and career length. Each cell contains the Pearson correlation value for the corresponding layer and metric.}
\label{table:pearson_perLayer}
\begin{tabular}{c c c c c }
\toprule
\multicolumn{2}{c}{}                                &  \textbf{Productivity}       & \textbf{Impact} &  \textbf{Mobility}             \\ 
\multicolumn{2}{c}{}                                & Number of publications   & H-index          & Number of institution changes \\ \midrule 
\multirow{5}{*}{\rotatebox[origin=c]{90}{\textbf{Layer}}} & 1 & 0.26                & 0.28    & 0.24                   \\ \cmidrule{2-5} 
                                                  & 2 & 0.51                & 0.50    & 0.45                   \\ \cmidrule{2-5} 
                                                  & 3 & 0.64                & 0.59    & 0.57                   \\ \cmidrule{2-5} 
                                                  & 4 & 0.73                & 0.64    & 0.64                   \\ \cmidrule{2-5} 
                                                  & 5 & 0.80                & 0.68    & 0.68                    \\ \bottomrule
\multirow{1}{*}{{\textbf{Career Length}}} &  & 0.49                & 0.57    & 0.37            \\ \bottomrule
\end{tabular}
\end{table}

  
Having established a clear association between productivity and ego network layer size, we next investigate how this relation is affected by the career stage and the mobility status (domestic vs international migrant). 
In Figure~\ref{fig:egonetwork_numpubs_cor_career_group}(a) we show, for each career stage, the Pearson correlation between the number of publications and the size of ego network layers. 
We plot the results for international migrants in blue and those for domestic ones in red. 
Please note that each line offers a compact summary of  the correlations extracted from a set of scatterplots equivalent to those in Figure~\ref{fig:egonetwork_pubs}. 
By and large, the trend observed in Figure~\ref{fig:egonetwork_pubs} globally is confirmed for individual career stages: researchers can easily grow their outermost collaboration circles (this stems from the high correlation for those layers), but establishing strong collaborations that affect the inner layers is costly, and aggressive publication is not sufficient to invert this trend, in any stage of career. 
It is interesting to note that international migrants tend to exhibit slightly stronger correlations  between circle sizes and productivity, but this difference is strongly significant (i.e, showing non-overlapping confidence intervals) only for associate and full professors. Considering the trends observed in Figure~\ref{fig:egonetwork_numpubs_cor_career_group}(a) up to the level of full professorship, results indicate that, eventually, researchers who migrate also internationally might be able to ``exploit'' more efficiently collaborations in their community, resulting in a higher correlation between layers sizes and number of publications. However, such a higher efficiency manifests itself from a certain stage of career onwards and the cumulative effect of international migration at the end of the career is not particularly significant.
We also observe that, typically, the longest the career duration, the highest the correlation for all the ego network layers.

 \begin{table}[t]
\centering
\caption{Tally of heavier publishers (corresponding to the top 10\% researchers from the productivity standpoint) in the different career and migration classes. A class is considered representative if there are at least 150 samples. We wrap within parentheses classes with too few samples.}
\label{table:top_people_prod}
\begin{tabular}{ c c c }
\toprule
\multirow{2}{*}{\textbf{Career Group}} & \multicolumn{2}{c}{\textbf{Heavy Publishers}} \\ \cmidrule(r){2-3} 
                                       & Int'l          & Domestic  \\ \midrule
Young Res                      & (98)                     & 232                        \\ 
Assistant Prof                   & 821                    & 839                        \\ 
Associate Prof                   & 3,192                   & 2,036                      \\ 
Full Prof                        & 586                    & 216                        \\ 
Distinguished Prof               & 228                    & (67)                         \\ \bottomrule
\end{tabular}
\end{table}

The final test that we carry out focuses on the most productive authors. We classify authors as most productive if they are in the 90th percentile of the productivity distribution. 
The number of samples in the most-productive group per class is reported in Table~\ref{table:top_people_prod}\footnote{Please note that the two columns do not add up exactly to the 10\% of authors in the corresponding career group (see Table~\ref{table:career_samples}). This is due to the fact that some authors have exactly the same number of publications, and we retain all of them for the analysis.}.
For young researchers and distinguished professors, there are just few samples of heavy publishers for certain mobility classes (e.g., only 67 distinguished professors in the domestic migrants class), hence these classes will be discarded from the analysis. 
The reason to analyse the most productive authors is the following. 
Results in Figure~\ref{fig:egonetwork_numpubs_cor_career_group}(a) have established a clear correlation between sizes of layers and productivity, particularly at the external layers, for the authors in our dataset. 
This seems to suggest that heavy publisher have extremely large ego co-authorship networks.
In order to investigate this claim, we compare the correlation observed for them (dashed lines in Figure~\ref{fig:egonetwork_numpubs_cor_career_group}(b)) with the correlation observed for the rest of the population (continuous lines). Recall (Section~\ref{sec:correlation_analysis}) that both correlations are adjusted with the Thorndike correction, to take into account the range restriction due to the 10\%-90\% split.
In general, we observe that the correlation for heavy publishers is significantly lower than for regular scientists, with the only exception of young researchers. This result is quite interesting, as it shows that heavy publishers do \emph{not} increase their ego network at the same pace of the general population, or, from the complementary standpoint, that nurturing very large collaboration networks does not necessarily lead to extremely high productivity. This can be interpreted as a \emph{capacity bound effect}: beyond a certain level, the most prolific researchers hit a capacity bound (in terms of number of collaborations) that limits the growth of their ego co-authorship network as the number of publications increase. This is a first result suggesting the presence of cognitive capacity bounds similar to the ones driving the structure of ego networks in general, also in the case of ego co-authorship networks. Note that the social capacity bound of Dunbar~\cite{dunbar1992time,roberts2009exploring} blends in cognitive limits (e.g., for remembering personal details about the alters and acting upon them) with temporal constraints (e.g., time spent grooming among primates). This might as well be the case for scientific collaborations, where both temporal and cognitive constraints are expected to play a role in the scientist's ability to keep up with many collaborators. An alternative explanation is that heavy publishers saturate the pool of potential co-authors, hence cannot grow their ego network layers at the same pace of regular researchers. While it seems unlikely that the finite number of potential collaborators plays a role here, we leave the exploration of this alternative hypothesis as future work.

\subsubsection{Research impact and collaboration circles} 
\label{sec:impact_and_ego}

Impact is the other dimension of performance, beside productivity, on which researchers are typically evaluated. Impact is usually measured by the h-index, which, despite its limitations~\cite{costas2007h-index}, is still by far the most widely used metric. In this section, \emph{we investigate whether high impact is associated with larger collaborations in the ego network circles}. Note that the h-index of an author is well-known to grow approximately linearly at the beginning of the career and to stabilise towards the end~\cite{hirsch2005index}.

 \begin{table}[t]
\centering
\caption{Tally of researcher with high impact (corresponding to the top 10\% researchers from the h-index standpoint) in the different career and migration classes. A class is considered representative if there are at least 150 samples. We wrap within parentheses classes with too few samples.}
\label{table:top_people_impact}
\begin{tabular}{ c c c }
\toprule
\multirow{2}{*}{\textbf{Career Group}} & \multicolumn{2}{c}{\textbf{Highest H-index}} \\
            \cmidrule(r){2-3} 
                                       & Int'l           & Domestic     \\ \midrule
Young Res                    &      (114)                  &        257                   \\ 
Assistant Prof                   &    915                    &     929                 \\ 
Associate Prof                   &     3,286                   &    1,941                \\ 
Full Prof                        &     595                   &     229                 \\ 
Distinguished Prof              &     221                   &    (79)                    \\ \bottomrule
\end{tabular}
\end{table}

Preliminarily, the overall correlation (with no class distinction) is reported in Table~\ref{table:pearson_perLayer} (second column). 
We observe that there is still a growing trend when moving from the innermost to the outermost layer. 
However, the correlation at the outermost circles remains lower than for the case of the number of publications.
Similarly to what we did for productivity, in Table~\ref{table:pearson_perLayer}, we also provide the correlation between the career length and impact. Again, such correlation is generally lower than the correlation between impact and layer size, hence we can rule out the hypothesis that the measured correlation is just a by-product of the career length effect.

Figure~\ref{fig:egonetwork_hindex_cor_career_group}(a) shows the correlation between h-index and circle size for domestic and international migrants (red solid line vs blue solid line) at different career stages. 
Similarly to the case of productivity, in Figure~\ref{fig:egonetwork_hindex_cor_career_group}(b) we also single out the top researchers in each category in terms of h-index (again, taking those above the 90th percentile) in order to study what happens for top performers (results shown as dashed lines). Recall that, for the 10\%-90\% split, we apply the Thorndike correction (Section~\ref{sec:correlation_analysis}).
Given that not enough samples are available for high-impact young researchers that are international migrants and distinguished professors that are domestic migrants (Table~\ref{table:top_people_impact}), we do not report their corresponding curves. 
We first focus on the general population of researcher (continuous lines, darkest colours) in Figure~\ref{fig:egonetwork_hindex_cor_career_group}(a). 
We note that, while the general trend is the same observed for the productivity, the absolute values of the correlation tend to be lower. 
More interestingly, while previously there were noticeable differences between domestic and international migrants, now these differences are less marked, with only a few cases where there difference is actually statistically significant (corresponding to non overlapping confidence intervals). In general, differently from the case of productivity, here correlation for international migrants is not greater  than for domestic migrants. This suggests that, while in terms of number of publications international migrants could exploit more efficiently their collaboration network, in terms of impact,  domestic and international migrants exploit collaboration networks in a similar way. This is a first result in our analysis showing that international migrations are not necessarily strategic in terms of publication performances.

If we distinguish between the general population and top researchers (from the h-index standpoint), the same properties can be observed in an even more evident way. Specifically, as it was the case for heavy publishers, we observe significantly lower correlation between h-index and co-authorship circles for the most impactful researchers. 
Indeed, Figure~\ref{fig:egonetwork_hindex_cor_career_group}(b) shows a somewhat counter-intuitive result, as one would expected very impactful researchers to grow their network easily, as they are very appealing as collaborators and probably sought after by many. 
Instead, contrary to this intuition, the collaboration circles of top researchers do not observe an outstanding growth. 
The career length seems to only slightly affect the absolute value of correlations, but the trends remain the same across all career stages. Moreover, this is another element suggesting that a capacity bound effect might be present: beyond a certain limit, having very big networks does not result in high performance, as there is not sufficient cognitive capacity to make those collaborations effective. 


\subsubsection{Take-home messages}
\label{sec:take_home_performance}

Summarising the results on academic performance and ego network structure, our findings are the following:
\begin{itemize}
    \item In general, there is a \emph{clear correlation between sizes of layers and performance indices} (both productivity and impact). Thus, as the productivity/impact grows, so does the collaboration circles. The correlation is particularly higher for external layers and this is in agreement with the theory on the different cognitive effort that is allocated to the collaborations in the different circles. The outermost circles only require low effort, while the innermost ones necessitate of intense nurturing, hence cannot be grown at will.
    \item \emph{International and domestic migrants show different efficiency levels in exploiting their collaboration networks}, respectively for productivity and impact. While international migrants seem more effective in exploiting larger networks for higher number of publications, this advantage is not present when considering high impact. This suggests that international migrations (which typically require more effort on the side of the migrating person) are not always particularly effective in terms of impactful research.
    \item The existence of a \emph{capacity bound effect} on the size seems confirmed: beyond a certain point (in terms of performance) higher performances are less and less correlated with large collaboration networks. This confirms at a higher level of detail the similarity between cognitive constraints in general social networks and in scientific collaborations. Also, this result seems to support the findings in~\cite{Petersen2015} about stable partnerships as a strategy for success.
\end{itemize}

\subsection{Mobility profile and co-authorship ego networks}
\label{sec:mobility_and_ego}

In the previous section, we have considered, among others, the impact of domestic and international mobility on the co-evolution of academic performance measures and collaboration circles. We now provide a more focused analysis on the interplay between co-authorship ego networks and the different dimensions of academic mobility. 

In Table~\ref{table:pearson_perLayer} (third column) we report, for each ego network layer, the correlation between its size and the number of affiliation changes. The trend is similar to that observed for the h-index, again showing a limited ability in growing the innermost circles. For the outermost circles (corresponding to weaker collaborations), the growth is significant, but still less than what we observe for the number of publications. For growing the weakest links, it seems, it is more important to publish extensively rather than striving for a high impact or moving a lot. Similarly to the case of productivity and impact, we can rule out the hypothesis that the measured correlation is just a by-product of the effect of career length, since its correlation with mobility is lower.

Figure~\ref{fig:egonetwork_movement_knownlocation_cor_career_group} presents the correlation between co-authorship layer size and number of affiliations, separating the results per career stages and type of academic migration. We observe that, for international migrants, there are two distinct career phases. The first corresponds to the beginning of the career (young researcher and assistant professor stages), for which we observe correlation but not a particularly high one, especially for the innermost circles. In the second career phase (from associate professor on), the correlation significantly increases in the outermost circles with respect to the previous phase, but overall we observe minimal changes across the career stages in this second phase. Thus, it seems that young scholars have more difficulties in growing their networks as they move across institutions, while senior researchers are very effective in capitalising movements. 
This might be due to the typically lower effort spent on individual papers by more established scientists with respect to younger researchers.
Interestingly, this dichotomy is not present for domestic migrants, for which the correlation, more or less steadily, increases as the career progresses. Comparing the two cases, we observe that, on a layer-by-layer basis, the correlation values at the end of the career are basically equivalent between international and domestic migrants. However, the former class enjoys a bump in correlation early in the career. Specifically, international migrants seem to capitalise mobility in terms of growing their network of collaborations already at the stage of associate professor at the same level (or even higher) as what domestic migrants achieve at the end of their career.

\begin{table}[t]
\centering
\caption{Tally of highly mobile researcher (corresponding to the top 10\% researchers from the mobility standpoint) in the different career and migration classes. A class is considered representative if there are at least 150 samples. We wrap within parentheses classes with too few samples.}
\label{table:top_people_mobility}
\begin{tabular}{ c c c}
\toprule
\multirow{2}{*}{\textbf{Career Group}} & \multicolumn{2}{c}{\textbf{Highly Mobile}} \\ 
\cmidrule(r){2-3} 
    & Int'l           & Domestic  \\ \midrule
Young Res                      &      176                &      235                   \\ 
Assistant Prof                 &     1,160                 &     821		 		\\ 
Associate Prof                & 3,749                   & 1,947                   \\ 
Full Prof                          &     648                 &        176       \\ 
Distinguished Prof          &       256               &         (42)             \\ \bottomrule
\end{tabular}
\end{table}

Similarly to what we did in Section~\ref{sec:productivity_and_ego} for heavy publishers and in Section~\ref{sec:impact_and_ego} for researchers with high impact, we now study separately very mobile scholars. 
The correlation between layer size and mobility for the most mobile authors are denoted in scales of red in Figure~\ref{fig:egonetwork_movement_knownlocation_HighLow}, for the others in scales of green. 
Again, we consider as the most mobile authors those above the 90th percentile (in turn, the remaining scholars are assigned to the group of regular authors).
Table~\ref{table:top_people_mobility} reports that we have a reasonable number of samples for all career stages with the exception of distinguished professors that migrate domestically, which we thus discard from the analysis.
In Figure~\ref{fig:egonetwork_movement_knownlocation_HighLow}, we observe different trends for international and domestic migrants. Let us first discuss the former (first five rows of Figure~\ref{fig:egonetwork_movement_knownlocation_HighLow}). Highly mobile young researchers that migrate internationally exhibit larger correlations than their less mobile counterpart. This difference is statistically significant in all the outermost layers. Hence, the collaboration networks of highly mobile young researchers tend to be larger in presence of international mobility. This relationship, though, wanes quickly as the career progresses: there is basically no difference between highly-mobile and regular assistant professors. The trend flips starting from associate professors: we observe smaller correlation for highly mobile scientists than for the others. Thus, highly mobile international migrants easily grow their (weaker) collaborations at the beginning of their career, but rapidly fall behind their less mobile counterparts as the career progresses.  
Let us now examine the case of domestic migrants. For them, the differences in correlation between the highly mobile and the rest of the researchers is statistically negligible in the vast majority of cases. 
Thus, there seems to be a dichotomic behaviour: for international migrants, high mobility correlates with the collaboration network size more than for regular scientists, but only the very beginning of the career. After this initial phase, the relationship between mobility and collaboration network size instead becomes weaker for the highly mobile scientists with respect to the others.
For domestic migrants, the ``early boost'' is not even present and highly mobile researchers tend to be almost indistinguishable from the others. 

The next question we tackle is whether highly mobile researchers are different from the others when considering the relation between their productivity/impact and the size of their collaboration network. Note that in this set of results we rely on the Thorndike Correction Case 3 (see Section~\ref{sec:correlation_analysis}).
We start by investigating whether publishing more as a highly-mobile researcher pays off in terms of collaborations with respect to the average researcher.
We know, from Figure~\ref{fig:egonetwork_numpubs_cor_career_group}, that productivity correlates with the size of the collaboration circles, especially in the outermost layers (weaker collaborations). We also observed that heavy publishers, however, exhibit lower correlation than their average counterparts, hinting at the existence of a capacity bound.  
Figure~\ref{fig:egonetwork_pubs_Bymovement_knownlocation_HighLow} shows how the correlation between ego co-authorship layer size and productivity varies at the different stages of career and for the different classes of mobility. 
In general, highly mobile researchers are statistically indistinguishable from the others, apart for distinguished professors that have migrated internationally. For them, the correlation of the highly mobile is even lower than for the other career stages.
Thus, high mobility does not seem to pay off in terms of productivity and collaboration circles.
International and domestic migrants are substantially equivalent from this standpoint. The only significant difference is that highly mobile domestic migrants exhibit slightly smaller correlation values than international ones. This might suggest that the benefit of international migrations carries over to highly mobile researchers. 
%
%
%
In Figure~\ref{fig:egonetwork_hindex_Bymovement_knownlocation_HighLow} we focus on the impact of highly mobile researchers and on how it correlates with the number of collaborations at the different layers. At the beginning of the career, highly mobile researchers are basically indistinguishable from the others, in both migration classes. However, starting from associated professors, being highly mobile entails having smaller collaboration circles as the impact increases, with respect to regular researchers. This differences is larger for international than for domestic migrants.
%
%
%
%
%
Putting all the results together (from Figures~\ref{fig:egonetwork_movement_knownlocation_HighLow}, \ref{fig:egonetwork_pubs_Bymovement_knownlocation_HighLow} and \ref{fig:egonetwork_hindex_Bymovement_knownlocation_HighLow}), and considering that the authors in the most mobile groups are the same for the analysis of both performance features, we find that most mobile authors are not able to capitalise their mobility neither in terms of productivity nor of impact, as for them having bigger ego network sizes is less correlated with high impact/number of publications with respect to regular authors.
The same conclusion can be also obtained by inspecting the correlation between the two performance indices, i.e., number of publications and h-index (Figure~\ref{fig:pubs_hindex_Bymovement_knownlocation_HighLow}). Specifically, comparing the most mobile researchers with the rest, it is clear that their correlation is similar or lower than that of average researchers. This confirms that high mobility does not seem to pay off in terms of research performance.


\subsubsection{Take-home messages}
\label{sec:take_home_mobility}

In this Section~\ref{sec:mobility_and_ego} we have focused on the interplay between the mobility profile and the ego network structure. The main findings are summarised below.

\begin{itemize}
	\item International migrants enjoy a boost in their co-authorship ego networks earlier in the career with respect to domestic migrants. The latter, however, are able to catch up and end their career at comparatively the same ego network levels.
	\item	 The most mobile authors do not publish more nor increase the number of their collaborations with respect to less mobile researchers. Moreover, they are not able to capitalise mobility in terms of impact, as for them having bigger ego network sizes is less correlated with high impact with respect to regular authors.
\end{itemize}

\section{Conclusion}
\label{conclusion}

Within the Science of Science framework, in this work we have investigated the interplay between academic performance, academic collaborations, and academic migrations. The analysis was carried out leveraging a dataset of geo-referenced publications from 81,500 authors at different stages of career, collected from Scopus. The collaboration network is computed from the perspective of individual authors, and is linked to results from anthropology on hierarchical social structures. Specifically, the ego network structure of the collaborations is obtained for each author in our dataset. The innermost layer of an ego network contains the strongest collaborations, while the weakest ones can be found in the outermost layers. Our results show that increasing productivity and impact of a researcher correlate with larger ego network layers, but, while the outermost layers can be easily grown, the innermost ones are much less sensitive to the measured academic performance. In addition, we observed a capacity bound, whereby the marginal utility of an increasing performance is less and less correlated with larger collaboration networks. From the mobility standpoint, we have observed that international migration correlates with higher productivity more than with higher impact. Also, international migrants seem to benefit from a collaboration boost at earlier stages of their career compared to domestic migrants. However, the latter are able to catch up and end their career at approximately the same levels of collaborations. Finally, we observed a lack of pay-off for very high mobility: its correlation with research impact is smaller than for regular authors.
 
\section*{List of abbreviations}
\label{abbreviations}
No abbreviations have been used in this manuscript.

\begin{backmatter}

\section*{Availability of data and materials}
The data that support the findings of this study are available from Scopus but restrictions apply to the availability of these data and so they are not publicly available. Data are however available from the authors upon reasonable request and with permission of Scopus. The list of scientists analysed in this study is available from the corresponding author on request.

\section*{Competing interests}
  The authors declare that they have no competing interests.
  
\section*{Funding}
This work was partially funded by the SoBigData++ and HumaneAI-Net project. The SoBigData++ project has received funding from the European Union's Horizon 2020 research and innovation programme under grant agreement No 871042. The HumaneAI-Net project has received funding from the European Union's Horizon 2020 research and innovation programme under grant agreement No 952026. 

The work of Pavlos Paraskevopoulos was supported by the ERCIM Alain Bensoussan Fellowship Program.


\section*{Author's contributions}
Designed the study: PP CB AP MC. Collected and processed the data: PP. Analyzed the data: PP CB AP MC. Wrote the paper: PP CB AP. All authors read and approved the final manuscript.
    

\section*{Authors' information}

PP was a postdoctoral researcher at IIT-CNR at the time of the study. CB is a permanent researcher at IIT-CNR. AP and MC are research directors are IIT-CNR. 

\bibliographystyle{bmc-mathphys} 
\bibliography{refs.bib}      



\setcounter{figure}{0}  

\begin{figure}[!htp]
\begin{center}
\iffigures\includegraphics[scale=0.4]{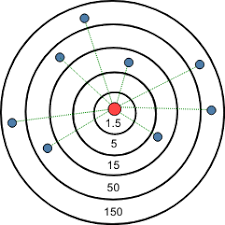}\fi
\caption{\csentence{Layered structure of the ego network.} The red dot at the center corresponds to the ego.}
\label{fig:egonet}
\end{center}
\end{figure}

\begin{figure}[!htp]
    \centering
    \iffigures\includegraphics[width=0.9\textwidth]{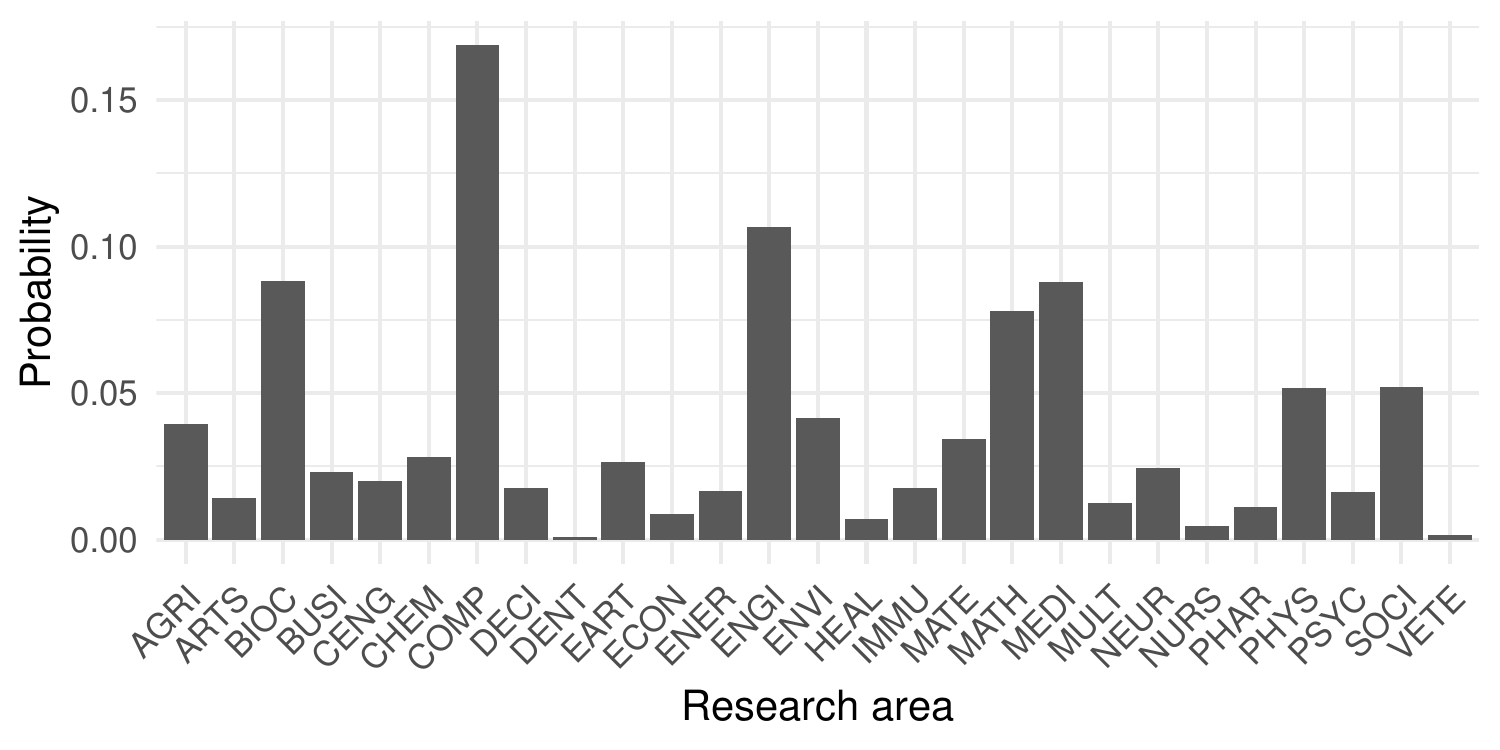}\fi
    \caption{\csentence{Distribution of the research areas covered by the authors in our dataset.} }
    \label{fig:histogram_authorsperarea}
\end{figure}

\begin{figure}[!htp]
    \centering
    \iffigures\includegraphics[width=0.9\textwidth]{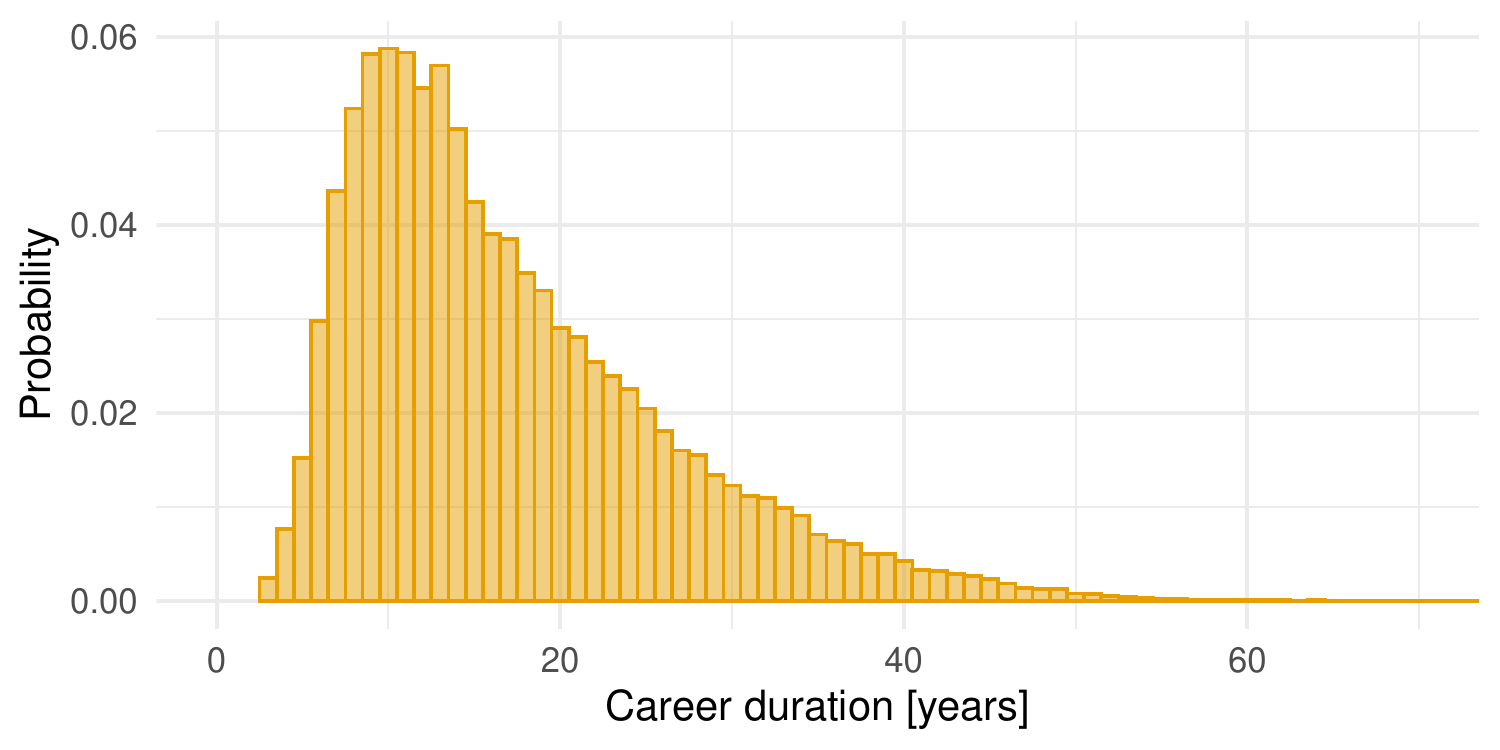}\fi
    \caption{\csentence{Career duration distribution.} The average career duration is 17.2 years.}
    \label{fig:histogram_careerduration}
\end{figure}

\begin{figure}[!htp]
    \centering
    \iffigures\includegraphics[width=0.9\textwidth]{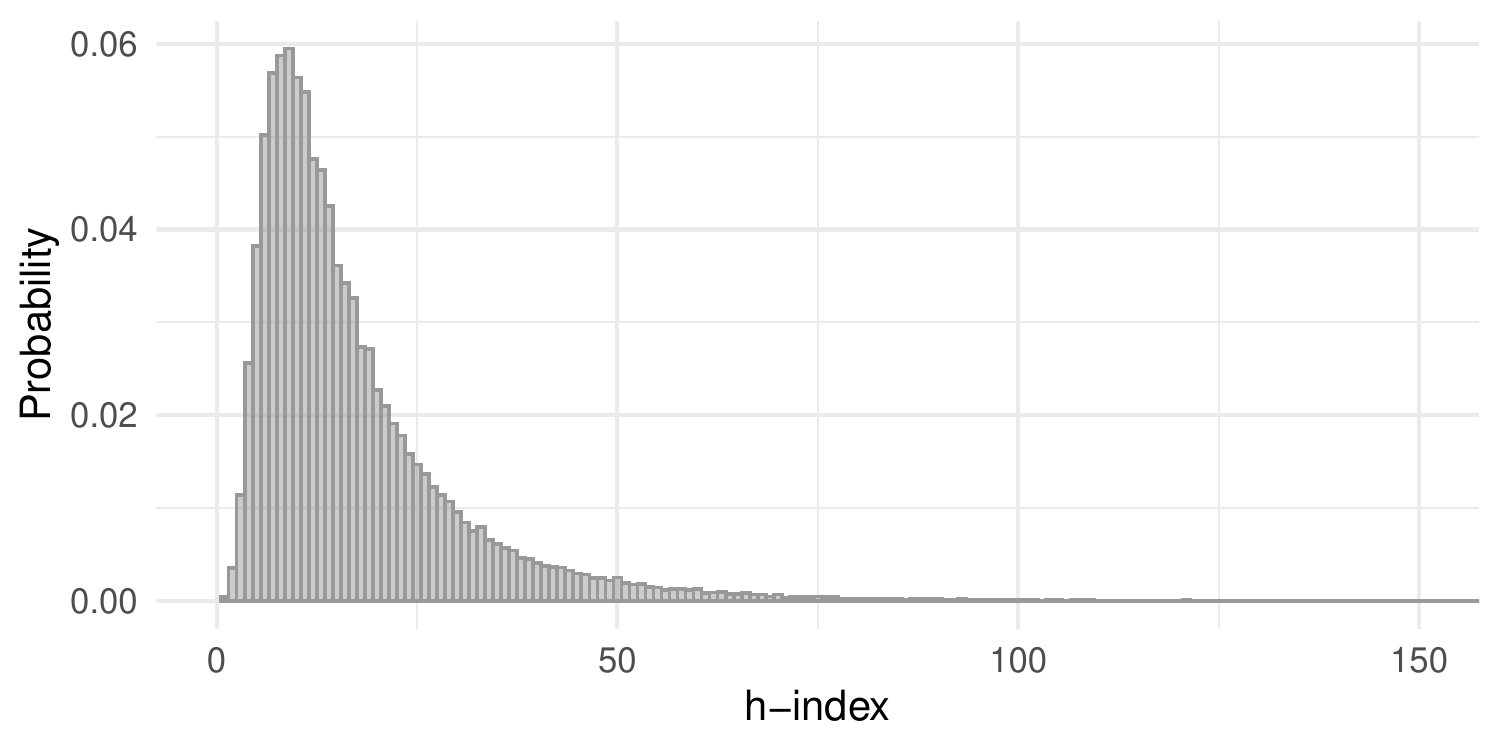}\fi
    \caption{\csentence{Impact distribution.} The average h-index for the authors in our dataset is 17.2.}
    \label{fig:histogram_hindex}
\end{figure}

\begin{figure}[!htp]
    \centering
    \iffigures\includegraphics[width=0.9\textwidth]{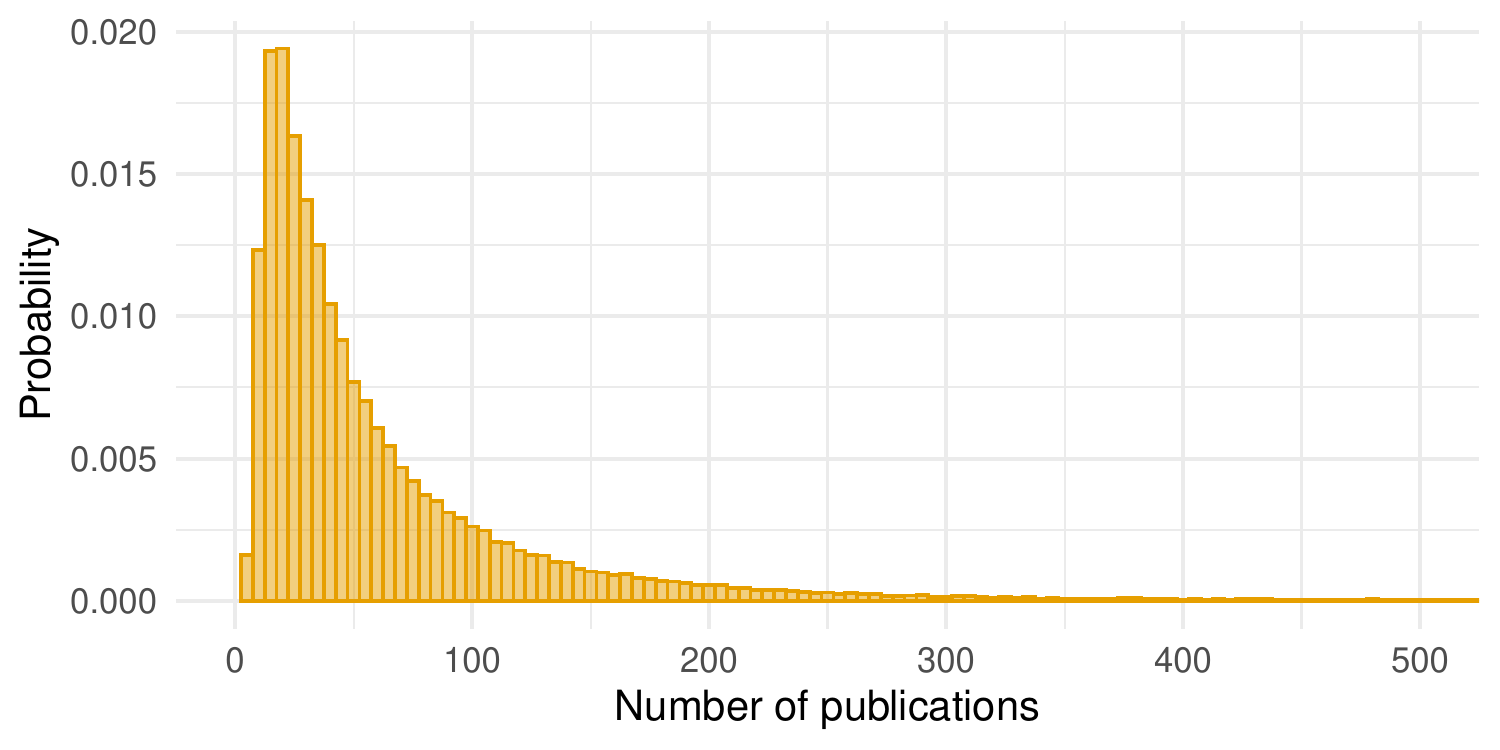}\fi
    \caption{\csentence{Productivity distribution.} The average number of papers published by the authors in our dataset is 65.7.}
    \label{fig:histogram_publications}
\end{figure}

\begin{figure}[!htp]
    \centering
    \iffigures\includegraphics[width=0.9\textwidth]{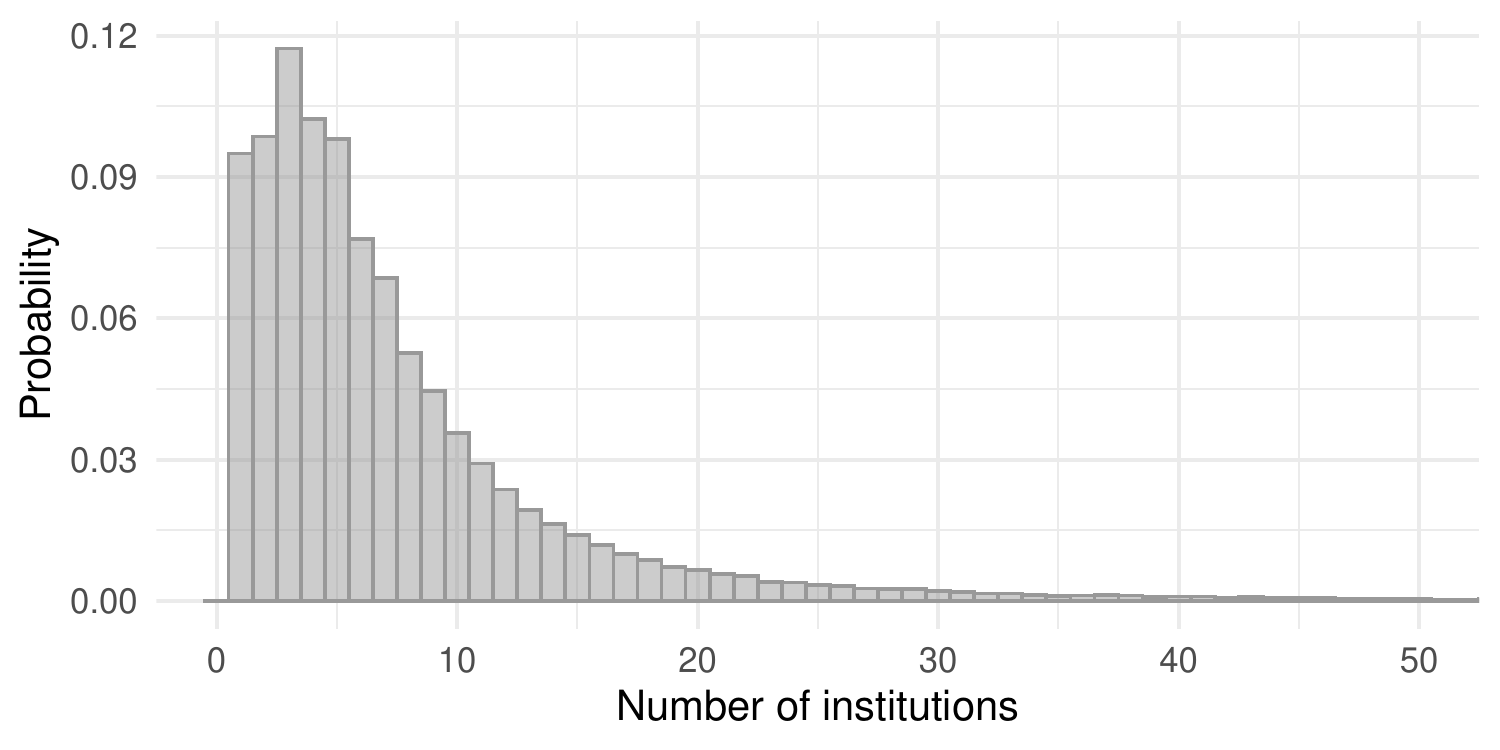}\fi
        \caption{\csentence{Distribution of the number of institutions changed.} On average, the researchers in our dataset have changed affiliation 7.8 times.}
    \label{fig:histogram_numberofmovements}
\end{figure}

\begin{figure}[!htp]
    \centering
    \iffigures\includegraphics[width=0.9\textwidth]{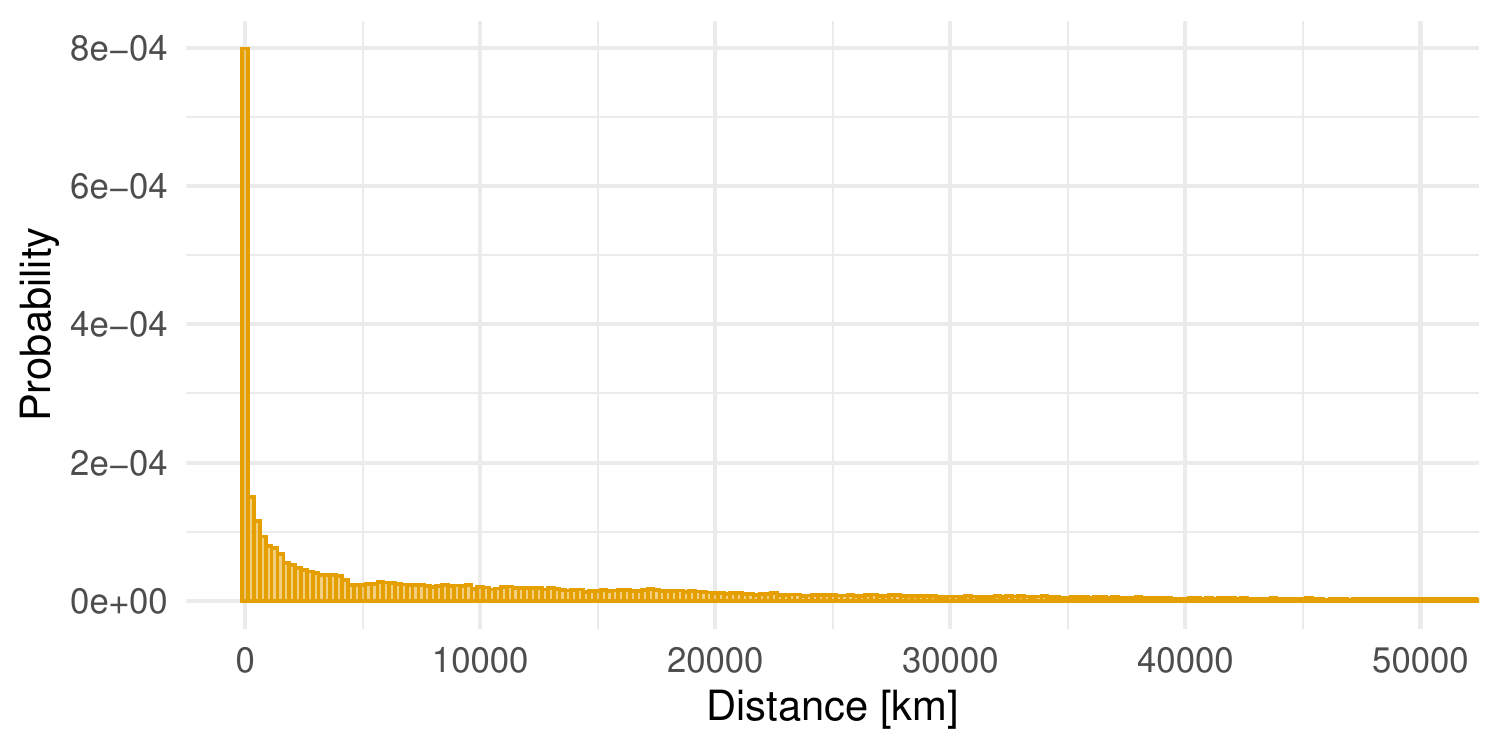}\fi
    \caption{\csentence{Distribution of the average distance travelled per trip.} }
    \label{fig:histogram_distofmovements}
\end{figure}

\begin{figure}[!htp]
    \centering
    \iffigures\includegraphics[width=0.9\textwidth]{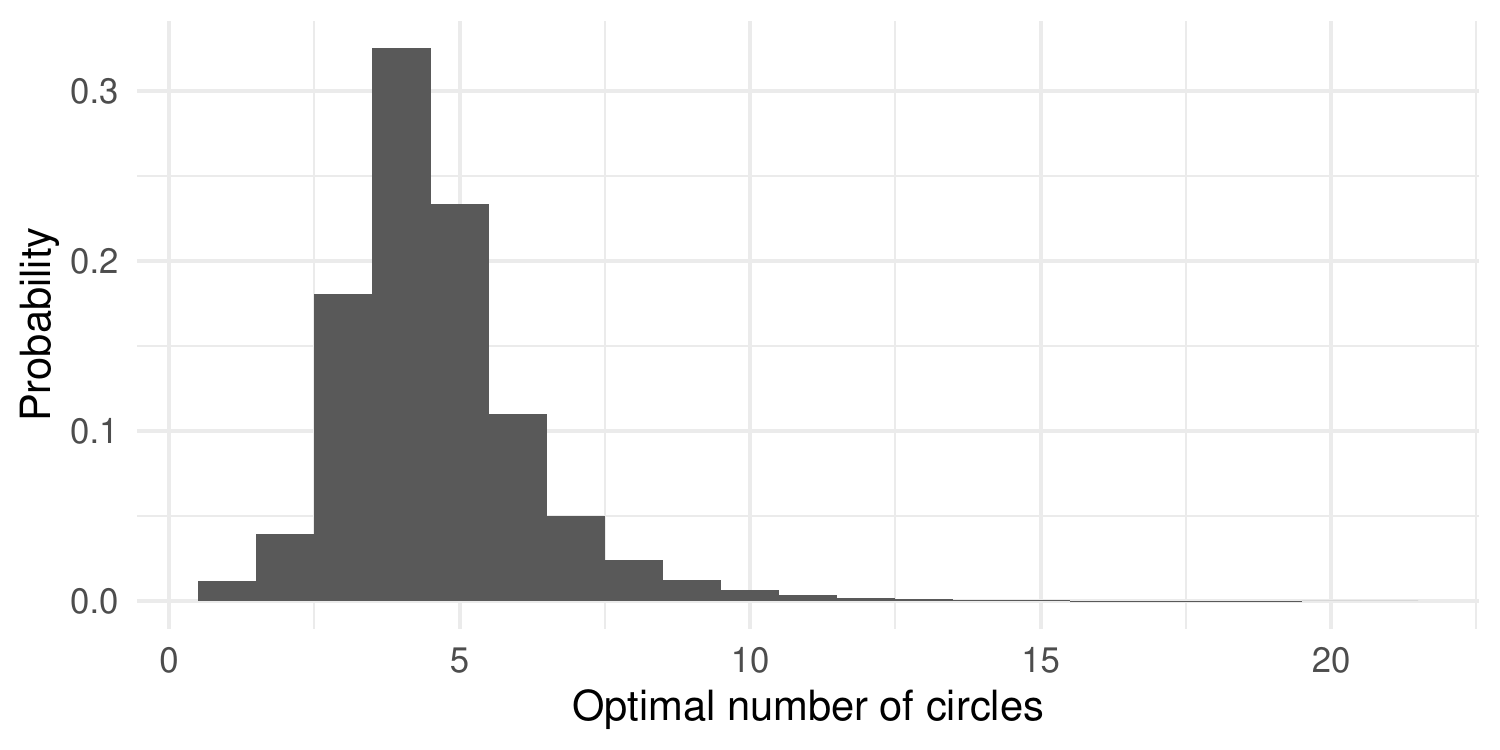}\fi
    \caption{\csentence{Distribution of the optimal number of collaboration circles.} }
    \label{fig:histogram_optimalcircles}
\end{figure}

 \begin{figure}[!htp]
    \iffigures\includegraphics[width=0.9\textwidth]{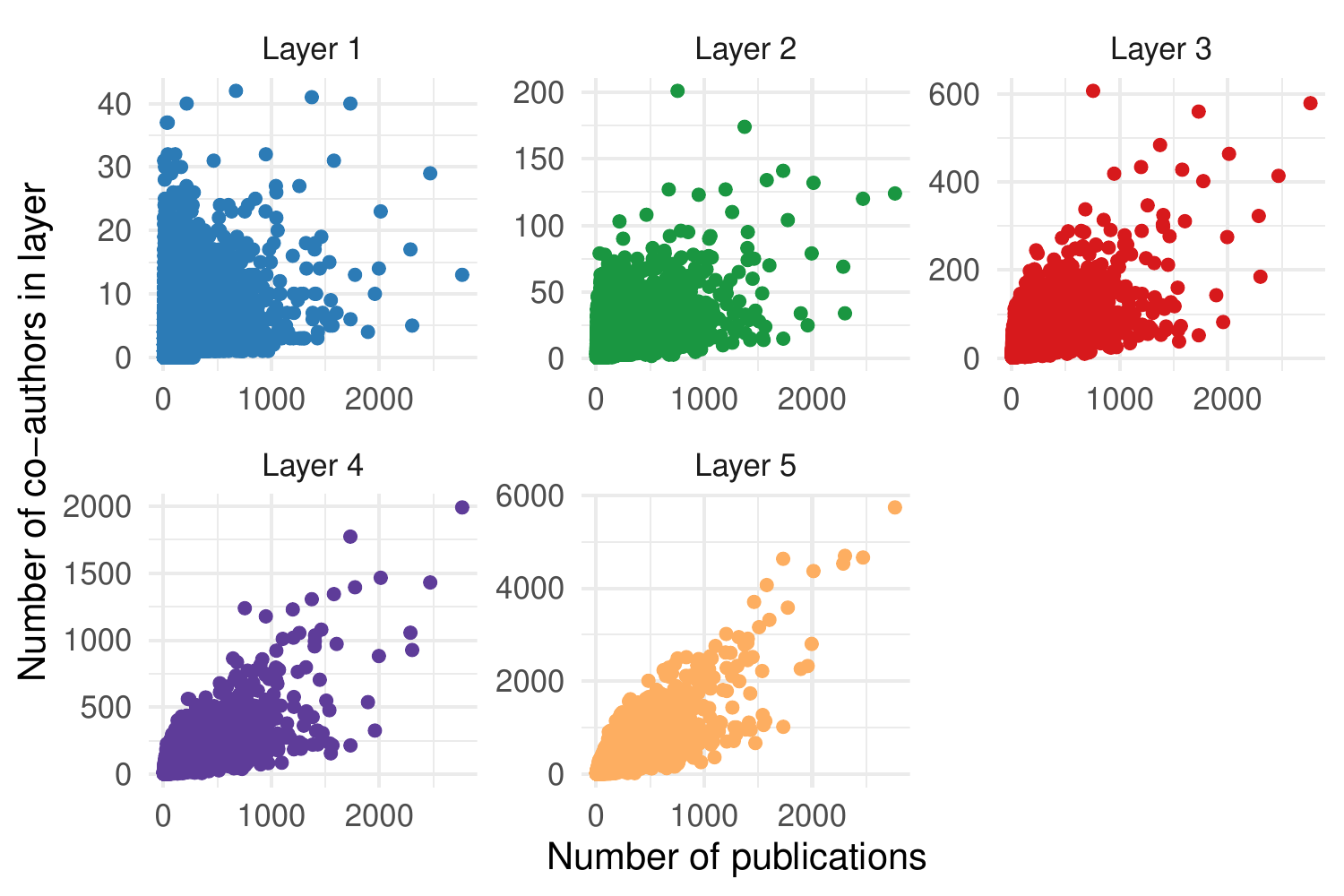}\fi
    \caption{\csentence{Productivity and co-authorship ego networks.}  Scatterplot of ego network layer size vs number of publications.}
    \label{fig:egonetwork_pubs}
\end{figure} 

 \begin{figure}[!htp]
    \iffigures\includegraphics[width=12cm]{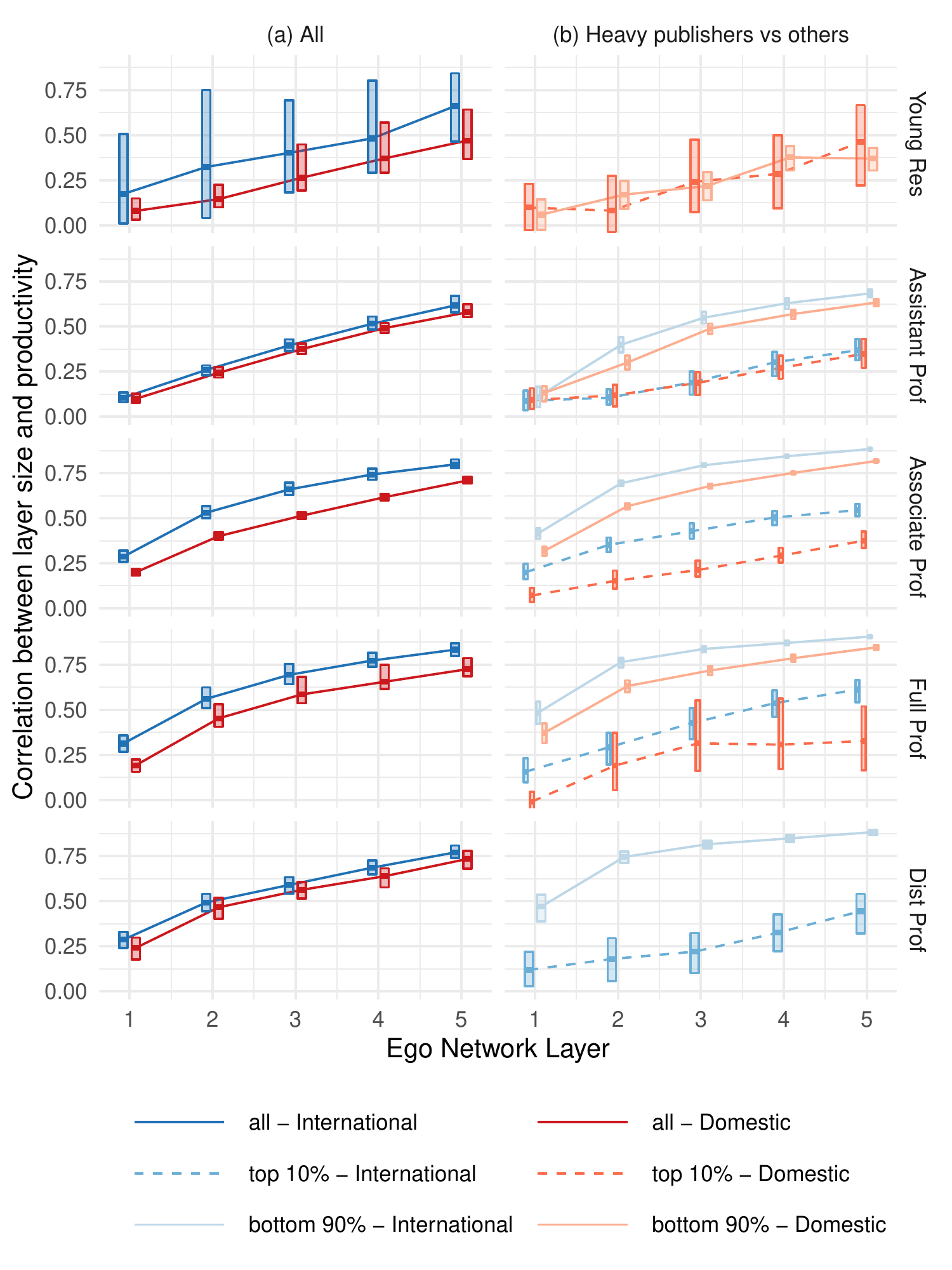}\fi
     \caption{\csentence{Productivity and co-authorship ego networks for international (blue) and domestic (red) migrants.} Correlation between ego network layer size and number of publications. Dashed lines are used for heavy publishers.}
     \label{fig:egonetwork_numpubs_cor_career_group}
 \end{figure}
 
 \begin{figure}[!htp]
    \iffigures\includegraphics[width=12cm]{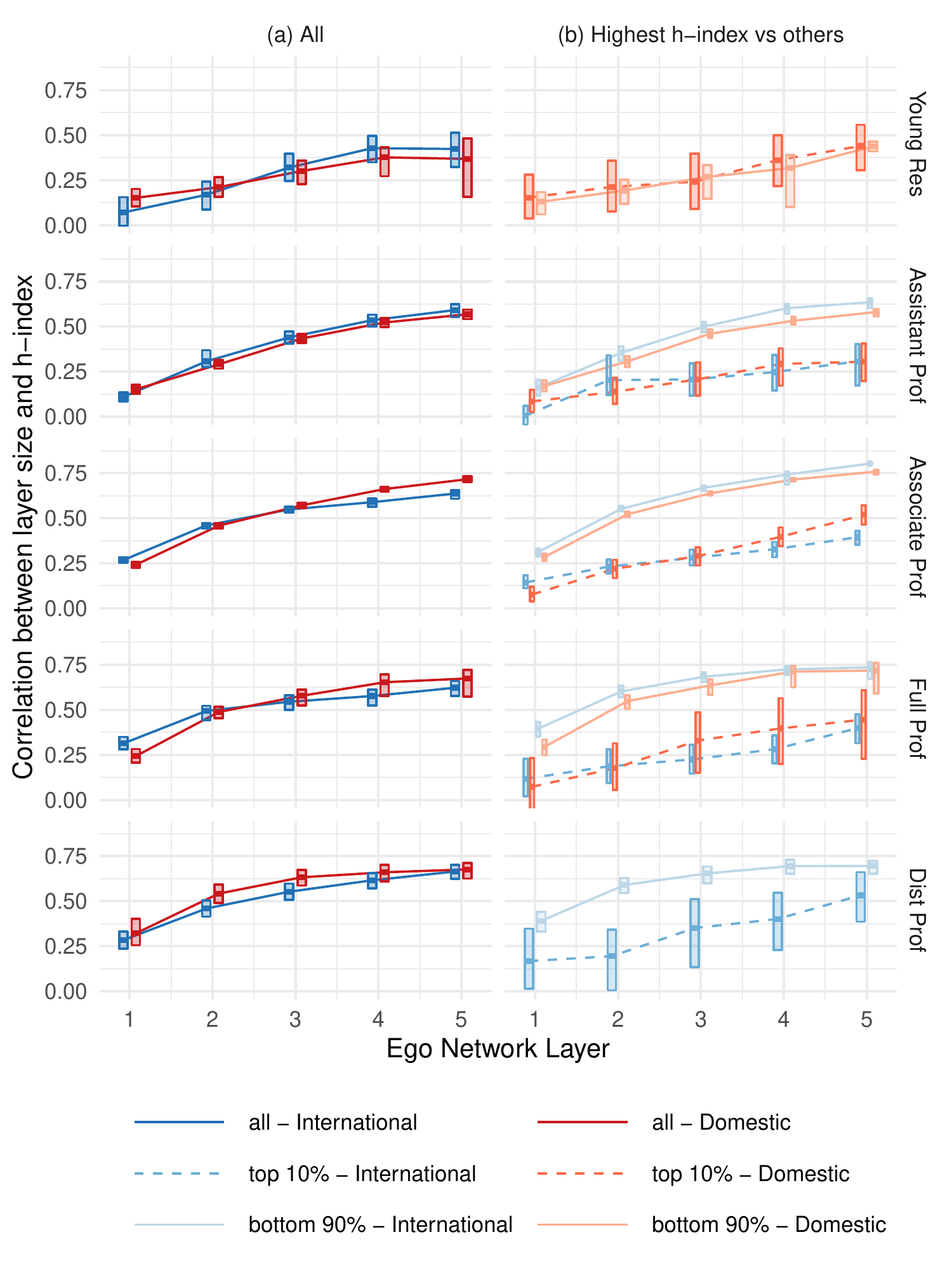}\fi
 \caption{\csentence{Impact and co-authorship ego networks for international (blue) and domestic (red) migrants.} Correlation between ego network layer size and the h-index.  Dashed lines are used for the group of researchers with the highest h-index.}
    \label{fig:egonetwork_hindex_cor_career_group}
\end{figure}

\begin{figure}[!htp]
    \iffigures\includegraphics[width=12cm]{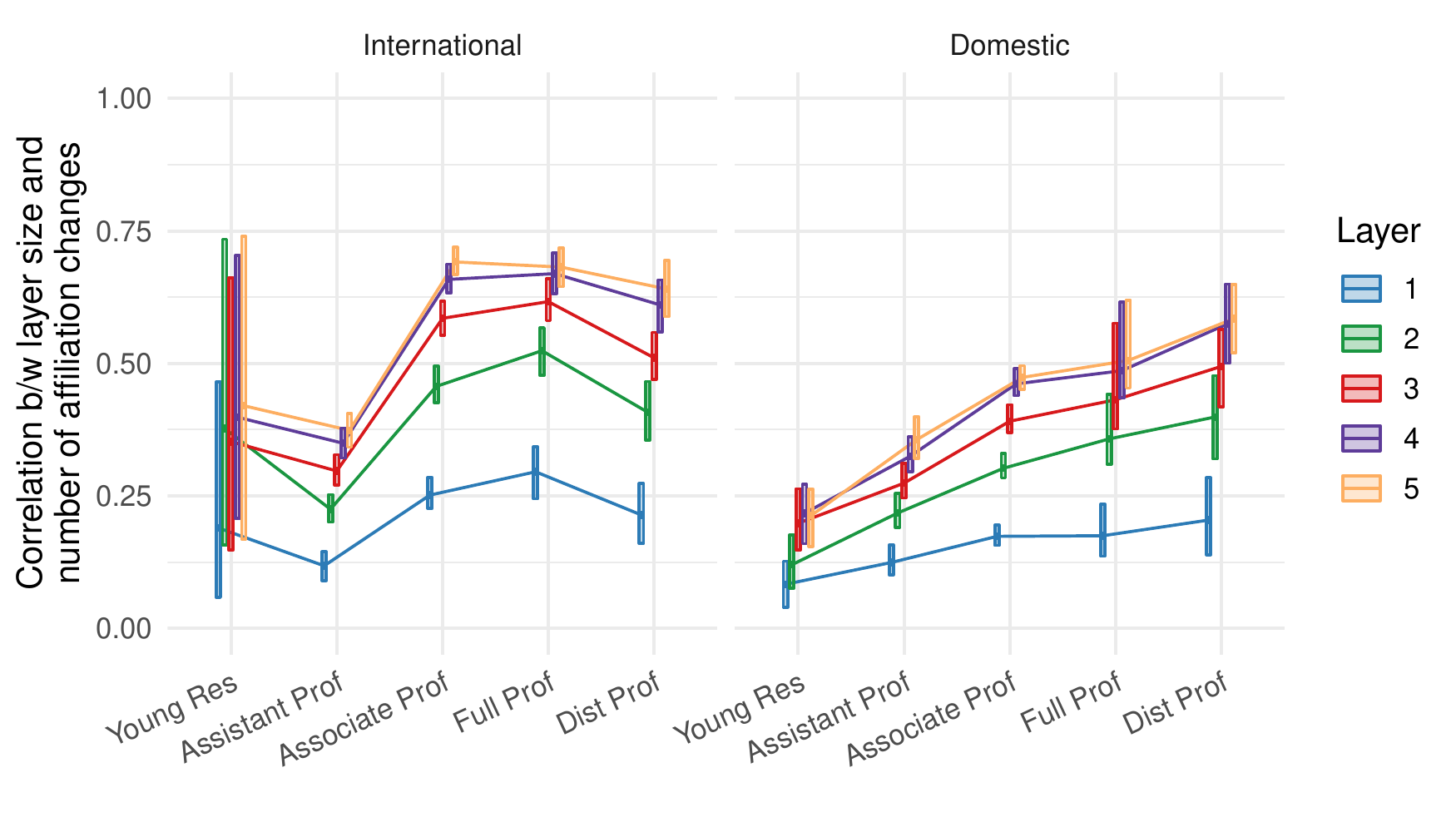}\fi
    \caption{\csentence{Mobility profile and co-authorship ego networks  for international and domestic migrants.} Pearson correlation between ego network layer size and number of institutions changed at the different career stages.}
\label{fig:egonetwork_movement_knownlocation_cor_career_group}
\end{figure}

\begin{figure}[!htp]
  \iffigures\includegraphics[width=12cm]{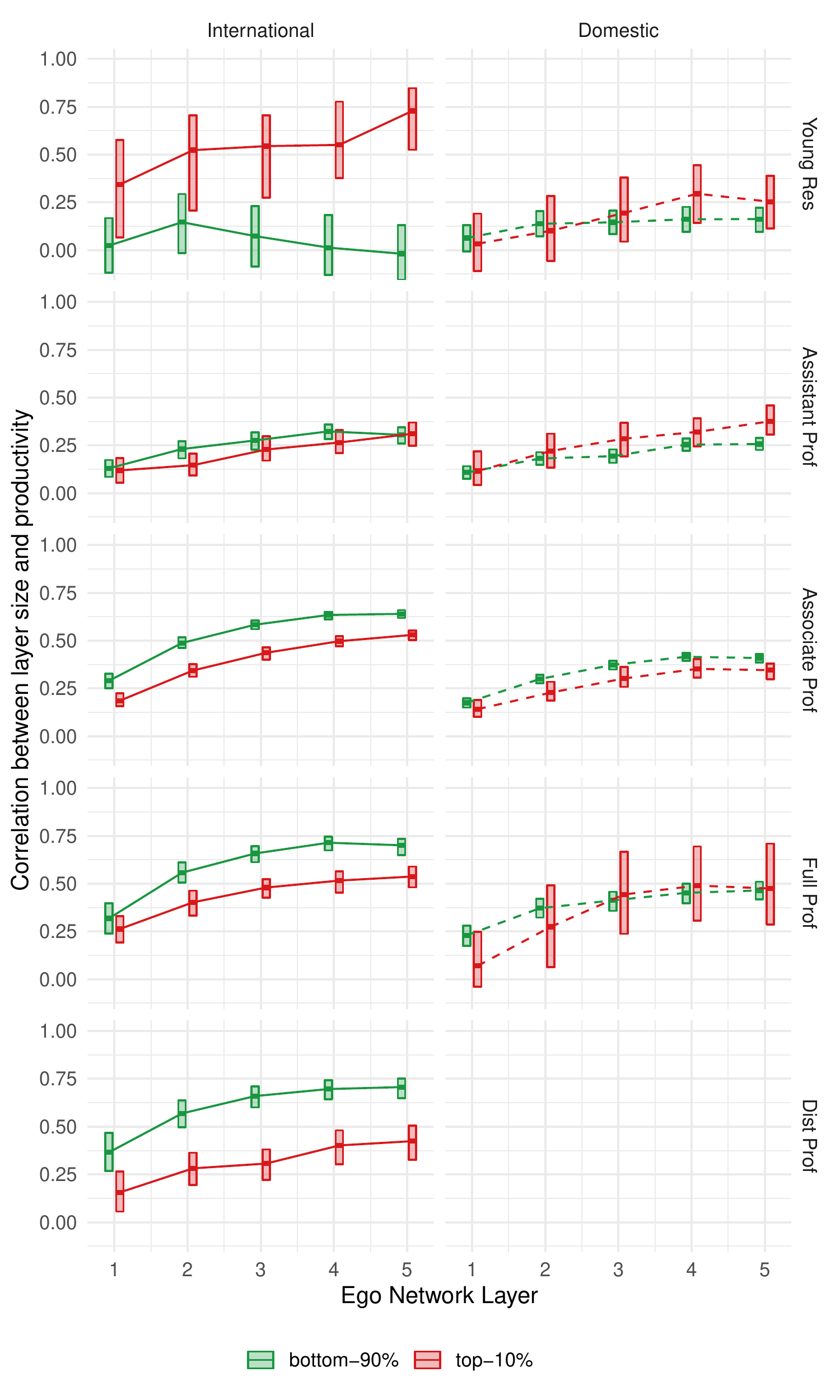}\fi
  \caption{\csentence{Mobility intensity and co-authorship ego networks.}  Correlation between ego network layer size and number of institutions changed, for highly mobile authors (red) and regular authors (green).}
     \label{fig:egonetwork_movement_knownlocation_HighLow}
\end{figure}

\begin{figure}[!htp]
    \iffigures\includegraphics[width=12cm]{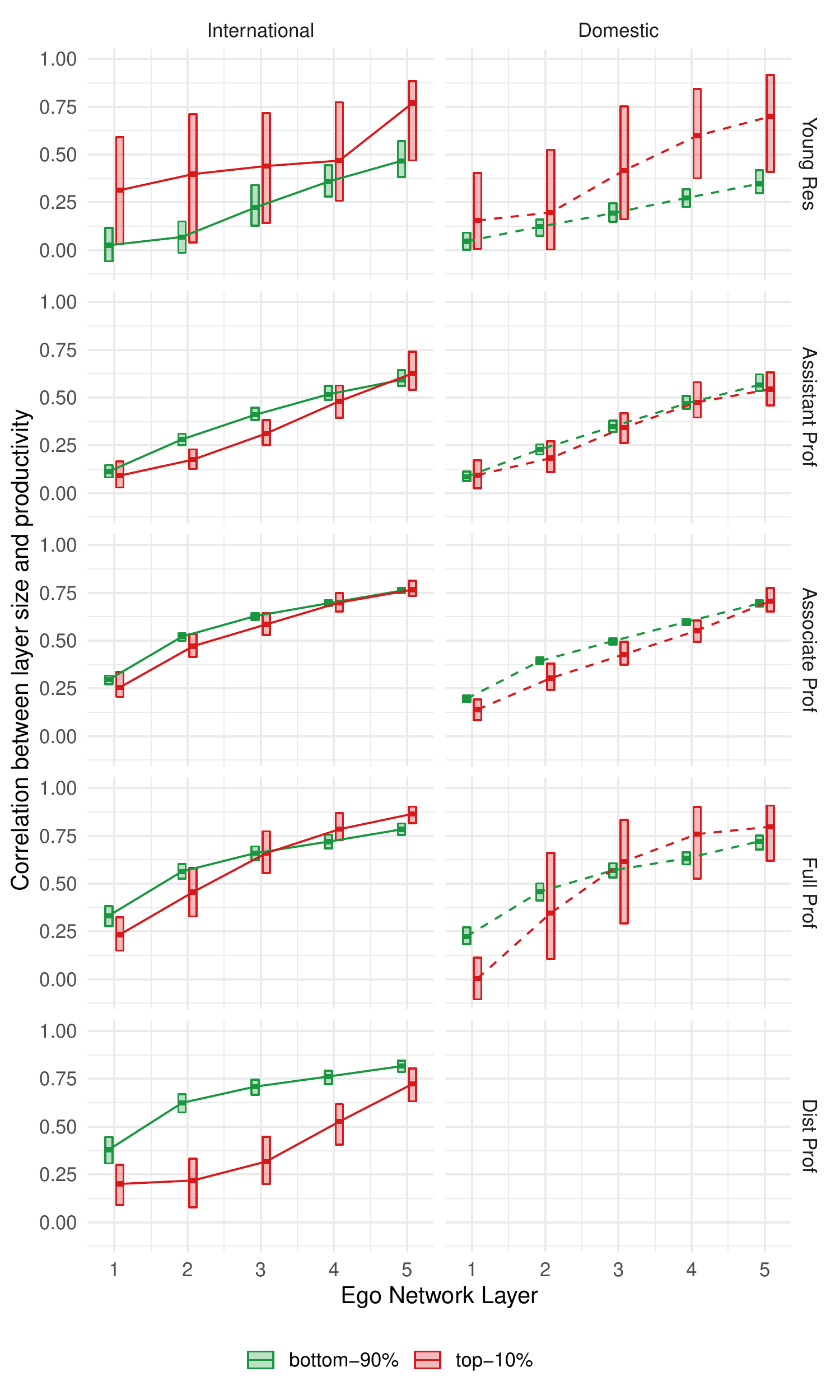}\fi
    \caption{\csentence{Highly-mobile researchers: productivity and co-authorship ego networks.} Correlation between ego network layer size and numbers of authored publications, for highly mobile authors (red) and regular authors (green).}
    \label{fig:egonetwork_pubs_Bymovement_knownlocation_HighLow}
\end{figure}

\begin{figure}[!htp]
    \iffigures\includegraphics[width=12cm]{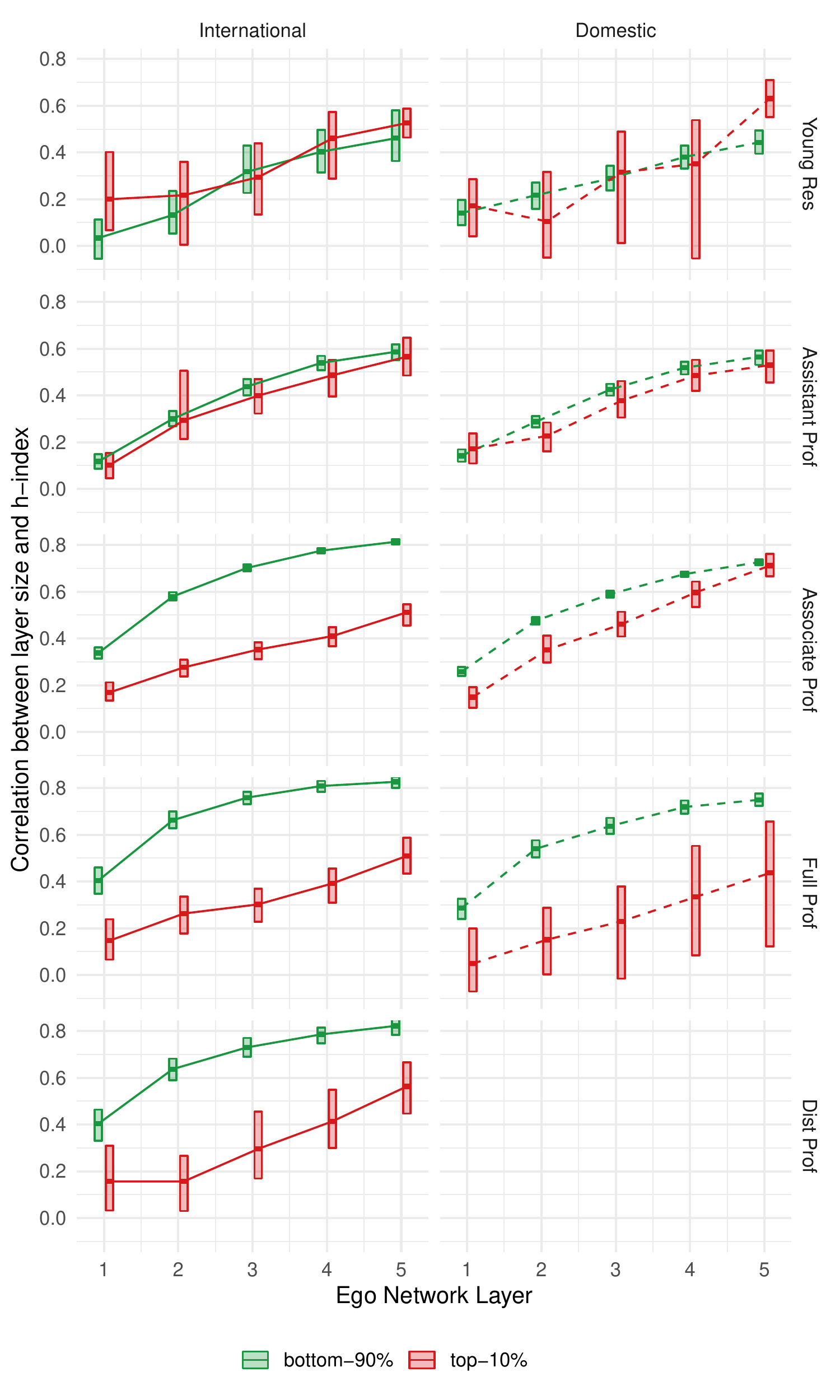}\fi
    \caption{\csentence{Highly-mobile researchers: impact and co-authorship ego networks.} Correlation between ego network layer size and h-index, for highly mobile authors (red) and regular authors (green).}
    \label{fig:egonetwork_hindex_Bymovement_knownlocation_HighLow}
\end{figure}

\begin{figure}[!htp]
    \iffigures\includegraphics[width=12cm]{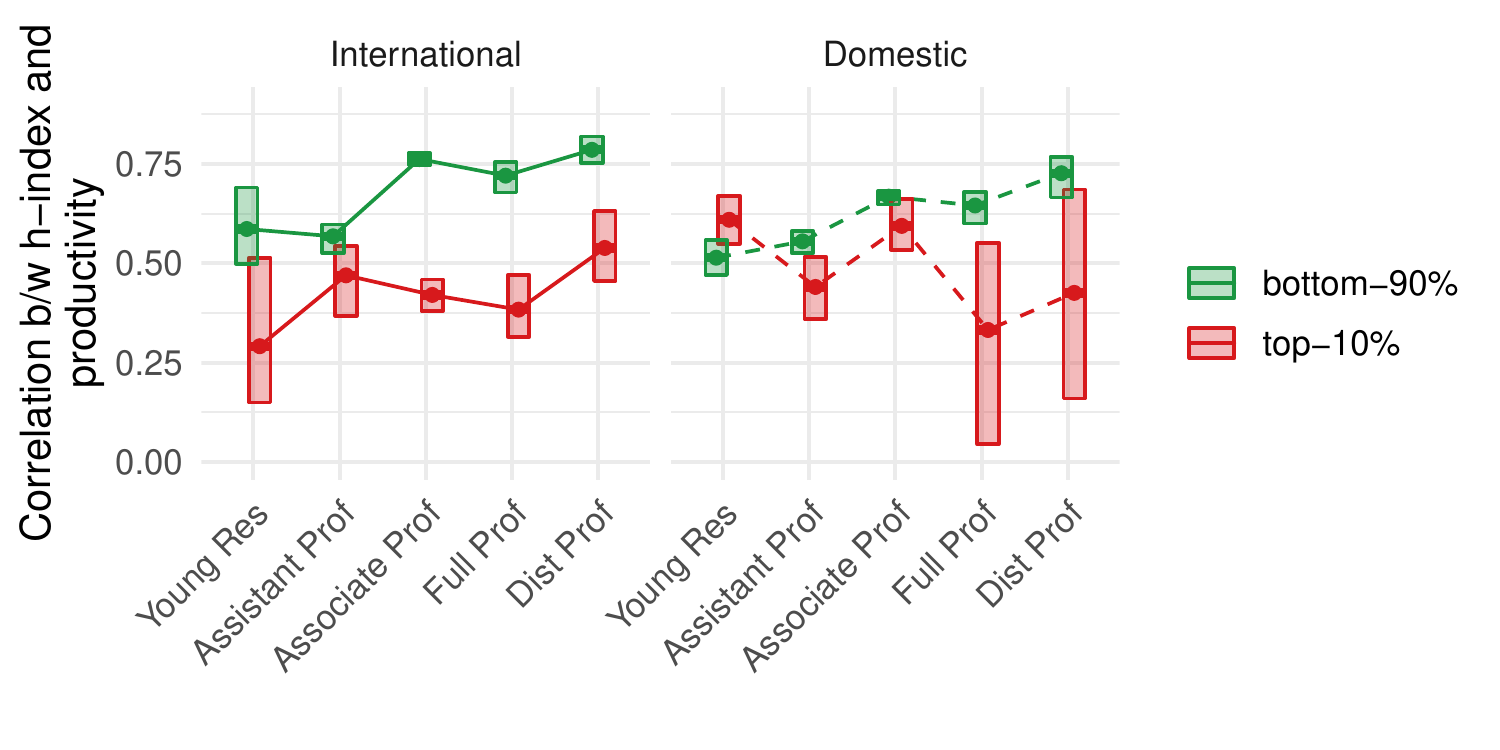}\fi
    \caption{\csentence{Highly-mobile researchers: productivity and impact.}  Correlation of publications and h-index by migration type and career stage. Highly mobile authors in red, regular authors in green.}
    \label{fig:pubs_hindex_Bymovement_knownlocation_HighLow}
\end{figure} 
\end{backmatter}
\end{document}